# EIGENRAYS IN 3D HETEROGENEOUS ANISOTROPIC MEDIA: PART I – KINEMATICS, VARIATIONAL FORMULATION

*Zvi Koren and Igor Ravve (corresponding author), Emerson*

zvi.koren@emerson.com , igor.ravve@emerson.com

## ABSTRACT

We present a new ray bending approach, referred to as the Eigenray method, for solving two-point boundary-value kinematic (Parts I, II, III) and dynamic (Parts IV, V, VI, VII) ray tracing problems in 3D smooth heterogeneous general anisotropic elastic media. The proposed Eigenray method is aimed to provide reliable stationary ray path solutions and their dynamic characteristics, in cases where the convergence to the stationary paths, based on conventional initial-value ray shooting methods, becomes challenging.

The kinematic ray tracing solution corresponds to the vanishing first traveltime variation, leading to a stationary path, and is governed by the nonlinear second-order Euler-Lagrange equation (Part I). In Part II we further elaborate on theoretical aspects of the proposed method and validate its correctness for general anisotropic media. within Part III we use a finite-element approach, applying the weak formulation that reduces the Euler-Lagrange second-order ordinary differential equation to the first-order weighted-residual nonlinear algebraic equation set. For the kinematic finite-element problem, the degrees of freedom are the discretized locations and directions along the ray trajectory. In Part IV, we propose an efficient method to compute the geometric spreading of the entire stationary ray path using the global traveltime Hessian, already established at the kinematic stage. This approach however is not a replacement for the solution of dynamic ray tracing , since it it solves only a limited dynamic problem and does not deliver the geometric spreading for intermediate points along the ray, nor the analysis of caustics.

In Part V we formulate the actual dynamic ray tracing, considering the Lagrangian-based second traveltime variation, which leads to the linear second-order Jacobi equation, and in Part VI we relate the proposed Lagrangian approach to the more commonly used Hamiltonian approach, applied to the dynamic ray tracing in isotropic and general anisotropic media. For the dynamic problem, the degrees of freedom are paraxial normal-shift vectors and their derivatives with respect to the arclength of the central ray, defining the corresponding ray tube geometry. The solution is provided in Part VII, where we naturally implement a similar finite element approach applied for the kinematic problem.

In both kinematic and dynamic problems, in between the nodes, the values of the ray characteristics are computed with the Hermite interpolation, which we find most natural when applying the proposed finite-element formulation and implementation, in particular, for anisotropic media.

We distinguish two types of stationary rays, delivering either a minimum or a saddle-point traveltime (due to caustics), where each type is a result of a minimization process with its own target function.

In this part, we formulate the second-order nonlinear Euler-Lagrange kinematic equation with an original arclength-related Lagrangian, which we find most convenient and efficient for our proposed fine-element solver.  We provide the Hamiltonian related to this Lagrangian through the Legendre transform, and the corresponding Lagrangian-based and Hamiltonian-based ray equations.

## INTRODUCTION

Two-point ray tracing in general 3D heterogeneous anisotropic media is one of the cornerstones for simulating the propagation of the high-frequency components of seismic body-waves between sources and receivers. It is an extremely challenging task. It has been mainly performed using the ray shooting method followed by numerical convergence schemes to reach the destination point. In this method, a fan of rays is first traced from a given starting point to the acquisition surface, where groups of rays arriving near each target location (e.g., a receiver) with similar take-off angles (slowness vectors) are used for the convergence process. By covering a wide range of take-off angles, multi-pathing stationary solutions can be found. For example, Bulant (1996, 2002) suggests an original ray shooting algorithm that makes it possible to find all two-point trajectories for general 3D layer-based smooth inhomogeneous isotropic media. However, in complex geological areas, characterized by heterogeneity of the elastic properties of the rocks, the convergence to a given location can be highly sensitive to small changes in the take-off angles, resulting in shadow zones which the numerically traced rays can barely penetrate. The ray bending methods can be considered a complementary optimization approach to the ray shooting, where the proposed Eigenray method is an extension to currently available ray bending solutions. It primarily attempts to fill in the above mentioned shadow zones using the following workflow: a) constructing initial guess (non-stationary) trajectories by interpolating/extrapolating from nearby traced rays (e.g., provided by the ray shooting method), b) updating the initial guess trajectories until it satisfies Fermat's principole of stationary time: An iterative optimization procedure for finding the (nearest) stationary ray path between the two fixed endpoints using the finite-element approach, and c) weighting (quantifying) the plausibility of the solution with a proposed complexity criterion based on the computed amplitude loss (normalized geometric spreading, discussed in Parts IV and V of this study). As mentioned, the Eigenray method is particularly attractive for areas that involve considerable isotropic/anisotropic velocity variations (e.g., between sediments and salt/basalt/carbonate rocks) or local velocity anomalies (e.g., gas clouds) smoothed along the transition zones. The method can be also extended for "blocky" models with sharp velocity discontinuities across the surface interfaces, and we discuss this option as well. The case of simulating "head waves" is an extreme example of Eigenrays providing controlled plausible solutions where conventional ray tracing methods become extremely challenging. Thomson (1989) suggested the use of a ray bending (correcting) solution to study head (grazing) waves, where the primary (incidence) and secondary (reflected/scattered) waves travel (graze) along a given reflector/refractor. We note that the term "head waves" is used here to describe general phenomena of rays, mainly traveling laterally along/below transition zones or across/in-between local velocity anomalies.

Depending on the subsurface model representation, the geometry of the acquisition system and the problem to be solved, such as seismic forward modeling, migration and inversion, several different ray-based approaches for simulating the propagation of high-frequency body waves in inhomogeneous isotropic and anisotropic elastic media have been studied and implemented; each has its own advantages and disadvantages. These characteristics have been documented, for example, by Leidenfrost et al. (1999), where they study six different methods, namely, finite-difference (FD) eikonal solvers, the graph method, wavefront construction (WFC) and a combined FD and Runge–Kutta method, for calculating seismic traveltimes from a point source to a regular subsurface grid. Additionally, Thurber and Kissling (2000), suggest a method for classifying different strategies for computing ray paths and travel times combining ray shooting, bending, perturbation and grid-based approaches. Other methods and strategies, or combinations of different methods, are also available and widely used.

Wavefront construction (WFC) is an attractive method for simulating ray propagation in complex areas while attempting to avoid artificial “shadow zones”. Vinje et al. (1993) published a pioneering study on the wavefront construction method, along with later works by Lambaré et al. (1996), Lucio et al. (1996), Ettrich and Gajewski (1996), Gibson (1999), Gjøystdal et al. (2002), and Lai et al. (2009). In this method, a fan of dense rays is simultaneously propagated in time, where at each time step a new wavefront is constructed. During the propagation, the wavefront is normally expanded and hence, new ray segments, normal to the current wavefront, are added, to ensure reliable representation (sampling) of the wavefront. However, while crossing complex velocity regions, the wavefront begins to split into several branches (triplications) resulting in deformed (non-topological) shapes (e.g., with cusps). The process of adding ray segments (normal to the non-topological wavefronts) becomes very challenging, and the accuracy of the constructed wavefront decreases. In extreme cases, a huge number of additional rays is required to fulfil the required accuracy, making this method very expensive. Note that the WFC method does not naturally deliver the exact stationary ray paths, which is the main goal of the present work. If the ray paths are not explicitly needed, the method can be efficiently used to compute dynamic parameters as well. The WFC can be used along with the ray shooting and ray bending techniques, for example, to compute initial conditions for ray shooting and boundary conditions for ray bending.

Paraxial ray tracing methods have been intensively used to solve two-point ray tracing problems in smooth heterogeneous velocity media (e.g., Beydoun and Keho, 1987; Virieux et al., 1988;

Farra et al., 1989, Gibson et al., 1991; Farra, 1993, Strahilevitz et al., 1998) and for the computation of dynamic properties (e.g., Popov and Pšenčík, 1978). In this technique, paraxial (nearby) rays are approximated (predicted) from given reference (central) traced rays using the first-order perturbation theory. These methods can also be used to interpolate traveltimes in the vicinity of the central rays (e.g., Bulant and Klimeš, 1999).

Gaussian beam summation (Popov, 1982; Červený, Popov and Pšenčík, 1982) or Maslov's methods (Chapman and Drummond, 1982; Thomson and Chapman, 1985; Huang et al., 1998) are alternative methods for the two-point ray tracing, with the advantage of overcoming the ray theory singularity problems related to the vanishing ray Jacobian while crossing caustics.

The approach suggested in this study belongs to the class of the ray bending optimization methods which have been extensively studied in the past, mainly for isotropic media or anisotropic media with high level of symmetry (e.g., transverse isotropy). An early ray tracing approach based on Fermat's principle has been suggested by Wesson (1971). Julian and Gubbins (1977) derived a boundary-value formulation of the ray tracing equations that can be solved iteratively (a ray bending approach), claiming that in some cases this method is more efficient than the ray shooting method. Smith et al. (1979) applied the ray bending technique for a full 3D velocity inversion. Pereyra et al. (1980) extended the ray bending technique to allow reflection and transmission through the interfaces. In a later study, Pereyra (1992) applied a combination of a fast-shooting algorithm and a multipoint boundary-value ray bending approach to obtain several source-receiver arrivals (multipathing), and to compute geometric spreading. The method was then extended for complex 3D geological models (Pereyra, 1996). Thomson and Gubbins (1982) used equally spaced nodes along the horizontal $x$ axis and applied a cubic spline interpolation to solve the boundary-value ray tracing problem with the ray bending method. They also computed geometric spreading in areas that involve velocity anomalies. Thomson (1983) solved the inverse problem for the velocity model parameters by applying the ray bending method to relate changes in the geometric spreading to the lateral velocity variations. Um and Thurber (1987) presented the ray trajectory as a set of points with a linear interpolation between them, where they adjusted the locations of these points to fit the kinematic ray tracing equations. Westwood and Vidmar (1987) applied the ray bending method to simulate the signals interacting with a layered ocean bottom. Waltham (1988) studied models consisting of constant velocity layers separated by curved interfaces and computed ray paths, whose traveltimes are stationary with respect to (wrt) changes in the ray/interface intersection points. Moser (1991) used this method to compute the traveltime between the source point and all points of a given network. Moser et al. (1992) improved the conventional ray bending approach by applying: a) gradient search methods, and b) interpolation by beta-splines between the nodes. Farra (1992) applied the Hamiltonian formulation with the propagator matrix to the ray bending approach. Shashidhar and Anand (1995) solved the problem of 3D Eigenray tracing in an ocean channel. Grechka and McMechan (1996) developed a 3D two-point ray-tracing technique based on Fermat's principle. The suggested method takes advantage of the global Chebyshev approximation of both the model and the curved rays, and makes it possible to find minima, maxima, and saddle points of traveltime. Cores et al. (2000) assumed a piecewise-linear ray path and presented the problem of tracing rays under Fermat's principle in 2D and 3D heterogeneous isotropic media. They applied biharmonic splines to model the reflector geometry and the velocity function, where they used the global spectral gradient technique for the optimization.

Ecoublet et al. (2002) suggested a 2D ray bending tomography (as an alternative to an initial-value ray shooting tomography) where the traveltime between the endpoints satisfies Fermat's principle. Bona and Slawinski (2003) demonstrated that Fermat's principle of stationary traveltime holds for general heterogeneous anisotropic media. Zhao et al. (2004) applied an irregular network "shortest path" method for high-performance seismic ray tracing. This approach was later extended by Zhou and Greenhalgh (2005) for anisotropic media. Rawlinson et al. (2008) considered a variety of schemes for tracking the kinematics of seismic waves in heterogeneous 3D structures, including a ray shooting method, a ray bending method and a combined approach, with a linear interpolation between the nodes. Kumar et al. (2004) and Casasanta et al. (2008) developed a two-point ray bending algorithm for vertical transversely isotropic (VTI) media for compressional waves, applying the group velocity approximation suggested by Byun et al. (1989). Wong (2010) extended the method for layered tilted transversely isotropic (TTI) media, where the degrees of freedom (DoF) were intersections of the ray with the layer interfaces. Bona et al. (2009) developed a strategy for two-point ray tracing, using a stochastic simulated annealing global search method. Sripanich and Fomel (2014) presented an efficient algorithm for two-point ray tracing in layered media by means of the ray bending method, where the ray paths are discretized at the intersection of the rays with the structure's interfaces, applying the global traveltime Hessian and the Newton method to find a stationary ray path. Cao et al. (2017) suggested a fast-marching method to compute the traveltime along an expanding wavefront using Fermat's principle in transversely isotropic media with vertical and tilted axes of symmetry, where the ray (group) velocity was approximated from the moveout equation. Wu at al. (2019) applied the shortest-path ray tracing adhering to Fermat's principle, in order to suppress the noise and improve the quality of pre-stack seismic data (nonlinear optimal stacking). Hovem and Dong (2019) applied the Eigenray method to compute the trajectories, reflection points and incidence angles for a large number of rays in sea water with the same source and receiver, and multiple reflections from the sea floor and from the water surface. We note that in special seismic acquisition surveys, such as vertical seismic profiles (VSP) or well-to-well surveys, the ray bending approach can be particularly attractive.

Recently, Koren and Ravve (2018a) demonstrated the power of applying a ray bending solution (referred to as the Eigenray method) by using a nonlinear spectral element method (Lagrange elements) to efficiently find accurate stationary ray paths in complex geological areas, characterized by smooth heterogeneous isotropic media. In a following abstract (Koren and Ravve, 2018b), the theory was extended to general anisotropic media, using Hermite finite elements, which makes it possible to naturally impose continuity (or discontinuity) conditions at the ray nodes for both locations and directions of the ray velocity. This abstract is heavily based on our study of the computation of spatial and directional, first and second, derivatives of the ray velocity (governing the corresponding derivatives of the traveltime) in 3D smooth heterogeneous general anisotropic elastic media (Ravve and Koren, 2019). Two later abstracts on the kinematic Eigenray (Koren and Ravve, 2020) and dynamic Eigenray (Ravve and Koren, 2020) briefly summarize the basic theoretical and implementation concepts described in detail in this series of papers. The primary objective of the first three parts of the study is to provide a full derivation of the results presented in the kinematic Eigenray abstract and to further elaborate on related theoretical and implementation aspects.

In Part I we propose an original arclength-related Lagrangian and we derive the corresponding Euler-Lagrange nonlinear, second-order ODE kinematic equation. We also provide the corresponding Hamiltonian (related through the Legendre transform) and the Hamiltonian-based first order kinematic ray tracing equations in heterogeneous general anisotropic media for the flow parameter arclength.

In Part II we review alternative Lagrangians and their related Hamiltonians for general anisotropy (e.g., Červený 2002a, 2002b) and discuss their relation to those proposed in this study. The notations and definitions of the different Lagrangians and Hamiltonians used in this study are listed in Tables 1 and 2. We further validate the derivations by comparing the values of the ray equations components for several anisotropic scenarios, using the proposed Lagrangian and two different Hamiltonian approaches.

In Part III we apply a finite element approach to establish the stationary ray path, nearest to a specified initial-guess trajectory, for 3D smooth heterogeneous general anisotropic media.

In Part IV, we solve a particular dynamic problem without explicitly performing the dynamic ray tracing system: We compute the geometric spreading of the entire ray path between the source and receiver by condensing the global (all-node) traveltime Hessian into the source-receiver traveltime Hessian.

In Parts V, VI and VII, we compute the dynamic parameters along the resolving stationary rays by explicitly performing the dynamic ray tracing (variational formulation, relation between the Lagrangian and Hamiltonian DRT approaches, and finite-element implementation).

## THE KINEMATIC EIGENRAY METHOD

According to Fermat's principle, the ray path between two fixed endpoints is the one that leads to a stationary time (normally, the least time, but may also be a saddle point time, normally due to caustics). In Parts I and II we formulate and validate the variational approach to solve the nonlinear stationary traveltime two-point kinematic ray tracing (KRT) problem. In Part III, starting from an initial (non-stationary) trajectory discretized with a set of nodes, we implement the solution using a finite element approach, successively (iteratively) refining the locations and directions of the ray trajectory at the nodes. If a number of different stationary rays co-exist between the given endpoints (multi-pathing), a different initial trajectory is applied to each solution (see numerical examples in Part III).

Initial trajectories using the ray shooting method

Overall, as a strategy for performing two-point ray tracing for conventional seismic acquisition surveys, we recommend starting with the ray shooting method, followed by numerical convergence techniques. In cases where the numerical convergence is difficult (the solution becomes very sensitive to fine changes in the take-off angles), the proposed Eigenray method can be used as an additional (complementary) attempt to obtain plausible solutions. In these cases, a (non-stationary) solution between the endpoints is first predicted by interpolation/extrapolation from nearby rays. For example, Levi et al. (2013) describes such an approach where a spherical Delaunay triangulation technique (Renka, 1997) is applied over the take-off ray directions at the source, only for those rays successfully arriving to the acquisition surface. The source (starting point) can be located either on the surface (e.g., simulating refraction or diving waves) or at the subsurface (e.g., simulating point diffraction rays). Depending on the complexity of the subsurface model, a given receiver location on the surface can be enclosed by several triangles (with different size and shape), each related to a different source-based direction (multi-arrivals). This process provides an efficient interpolation technique for obtaining two-point (normally non-stationary) trajectories for each arrival solution.

For other acquisition geometries, such as well-to-well or vertical seismic profiling (VSP), other approaches for obtaining the initial guess trajectories can be used.

The proposed arclength-related Lagrangian

In this part (Part I), we apply the Euler-Lagrange formula over an integrand of the traveltime functional (Lagrangian) valid for smooth heterogeneous general anisotropic elastic media. We propose a specific Lagrangian with a clear physical interpretation, which we find most convenient and efficient for our proposed finite element solution for stationary rays (and for computing dynamic properties) in general anisotropic media.

For a given trajectory between two fixed endpoints (a stationary ray path or a non-stationary approximation path in its vicinity), the Lagrangian $L = L\left[\mathbf{x}(s), \dot{\mathbf{x}}(s)\right]$ used in this work is a function of the arclength-dependent location vector, $\mathbf{x}(s)$, and the arclength derivative of the location vector, $\dot{\mathbf{x}}(s) = d\mathbf{x}/ds$, along the trajectory. The arclength derivative of the location vector, $\dot{\mathbf{x}}(s)$ , noted by $\mathbf{r} \equiv \dot{\mathbf{x}}(s)$, represents a vector tangent to the trajectory at any point along the ray, normalized to the unit length, $\mathbf{r} \cdot \mathbf{r} = 1$, and hence it is the ray direction (or equivalently, the ray velocity direction) vector, $\mathbf{r}(s)$, at $\mathbf{x}(s)$. The condition for the trajectory to be a stationary traveltime ray path can be symbolically written as $t = \int_S^R L\left[\mathbf{x}(s), \dot{\mathbf{x}}(s)\right] ds \rightarrow$ stationary , where $S$ and $R$ are the fixed endpoints of the path. The solution for this type of integral problem is heavily based on the computation of the spatial and directional derivatives of the ray velocity (Ravve and Koren, 2019), which are the core computational components for both, converging to the stationary ray (kinematics) and computing the dynamic properties.

Discussion on the approximated stationary paths

Obviously, for the actual traveltime stationary solution, $\mathbf{x}(s)$ and $\dot{\mathbf{x}}(s)$ are the location and direction vectors of the physical ray path (note that the derivative $\dot{\mathbf{x}}(s)$ is also the ray velocity direction). However, vectors $\mathbf{x}(s)$ and $\dot{\mathbf{x}}(s)$ can also be considered location and direction vectors for nearby, non-stationary, trajectories (approximations of the ray path). Indeed, for the approximated paths, they are not the physical location and direction vectors of the actual stationary ray between $S$ and $R$; however, they still represent local position and direction for a short interval of another, physical ray in the proximity of the given point. This, in turn, means that the ray velocity magnitude (and all its first and second spatial and directional derivatives) can be computed along the approximated ray path as well. The computed physical characteristics are valid for any small portion of the approximated ray.

For the proposed Lagrangian (traveltime integrand), we apply the Euler-Lagrange equation that yields a nonlinear, second-order ordinary differential equation (ODE) for the ray locations and directions, in terms of the ray velocity, its gradients and Hessians. In Part II we validate the proposed Lagrangian, and in Part III we describe the finite element solution for the Euler-Lagrange kinematic equation.

Spatial and directional derivatives of the ray velocity

The local gradients and Hessians of the traveltime require the corresponding derivatives of the ray velocity along the local segments of the path (finite elements). Due to anisotropy, we deal with two types of ray velocity gradient vectors: spatial and directional, and three types of ray velocity Hessian matrices: spatial, directional and mixed. The method for computing the ray velocity derivatives is a core component of this study. It has been recently published by Ravve and Koren (2019) (a summary of the method is given in Appendix D) for general anisotropic (triclinic) media, where the higher anisotropic symmetries were considered particular cases. The method establishes the spatial and directional gradients, $\nabla_{\mathbf{x}} v_{\text{ray}}$ and $\nabla_{\mathbf{r}} v_{\text{ray}}$, and the spatial, directional and mixed Hessians, $\nabla_{\mathbf{x}} \nabla_{\mathbf{x}} v_{\text{ray}}$, $\nabla_{\mathbf{r}} \nabla_{\mathbf{r}} v_{\text{ray}}$ and $\nabla_{\mathbf{x}} \nabla_{\mathbf{r}} v_{\text{ray}} = \left( \nabla_{\mathbf{r}} \nabla_{\mathbf{x}} v_{\text{ray}} \right)^T$, where $\mathbf{x}$ and $\mathbf{r} = \dot{\mathbf{x}} = d\mathbf{x} / ds$ are the vectors containing the spatial coordinates and direction components of the ray velocity, respectively, and $s$ is the arclength along the ray.

Directional Derivatives: To avoid confusion, we note that through all parts of the paper, the term "directional derivative" has a special meaning. Unlike its common use, it does not mean a spatial derivative in a definite direction, like, $df / dn = \nabla_{\mathbf{x}} f \cdot \mathbf{n}$, where $\nabla_{\mathbf{x}} f$ is the spatial gradient of a scalar function $f$ and $\mathbf{n}$ is the given direction. By "directional derivatives", we mean derivatives (of the Lagrangian or those of the ray velocity magnitude) wrt the components of the ray direction (or ray velocity direction) $\mathbf{r}$. The directional derivatives of the ray velocity are not trivial and they are essential for understanding the correctness of the proposed Lagrangian and its implementation in the kinematic and dynamic Eigenray method (see Appendix E).

The input for the computation of the first and second spatial, directional, and mixed derivatives of the ray velocity magnitude, for a given point location $\mathbf{x}$ and a given ray direction $\mathbf{r}$, includes the anisotropic elastic properties of the medium $\mathbf{C}(\mathbf{x})$, and their spatial first and second derivatives.

Discussion on Shear waves

The governing relationships in all parts of this paper are valid for any wave mode in general anisotropic elastic media. However, it is well known that for this type of medium, ray tracing is more challenging for shear waves than for compressional waves. In particular, this is the case when using the (boundary-value) ray bending (the proposed Eigenray method), where the input (at each stage of the iterative solution process) involves approximated ray locations and directions at specific nodes. The complexity is due to the phenomena of multi-valued ray (group) surfaces: For any given direction of a shear-wave, there are multiple corresponding slowness vectors and ray velocity magnitudes (e.g., Grechka, 2017). Hence, if there is no additional information (e.g., the slowness vector at the source), the Eigenray method faces the multiplicity of all the available solutions related to the obtained stationary ray (the corresponding multiple slowness vectors and ray velocity magnitudes; hence traveltimes). However, this type of complexity also arises in the (initial-value) ray shooting method, where the same ray path can be obtained with multiple slowness vectors at the source. Recall, that our recommended workflow suggests starting with the ray shooting method, and using the Eigenray method for the final convergence. We further emphasize that this challenge is mainly related to the kinematic ray tracing (KRT) stage, where the stationary path is obtained for a specific solution branch. In the dynamic ray tracing (DRT) stage, the resolving paraxial rays are in the infinitesimal proximity of the (central) stationary ray and represent a linearized solution. It is assumed that the paraxial rays share the same solution branch as the central ray.

In this study we only focus on compressional waves. Nevertheless, as noted by Červený (2002a, page 579; 2002b, page 223), Fermat's variational principle, with the use of the Lagrangian, can still be applied to shear waves in anisotropic media. The multi-valued ray velocity surface can be decomposed into several single-valued branches. Thus, one can choose a specific branch and proceed for that branch only.

Qualification using a complexity criterion

The Eigenray method can then be applied to adjust the interpolated/extrapolated initial-guess paths in order to obtain actual stationary ray solutions (if they exist). In Parts IV and V of this study we suggest using the dynamic characteristics of the Eigenray solution to define (compute) a "complexity criterion" which can qualify the plausibility of the obtained kinematic solution.

Appendices

In Appendix A, we prove that the generalized momentum related to our proposed arclength-related Lagrangian $L\left[\mathbf{x}(s),\mathbf{r}(s)\right]$ is indeed the slowness vector, $\partial L / \partial \mathbf{r} \equiv L_{\mathbf{r}} = \mathbf{p}$.

In Appendix B, we apply the Euler-Lagrange approach to the (stationary) traveltime integral, in order to obtain the second-order kinematic ray tracing ordinary differential equation (ODE) in terms of the ray velocity, its gradients and Hessians.

In Appendix C, we provide the Hamiltonian that correspond to the proposed arclength-related Lagrangian via the Legendre transform. We then provide the Hamiltonian-based ray tracing equations.

In Appendix D we elaborate on the computation of the ray velocity magnitude, given the medium elastic properties and the ray velocity direction.

In Appendix E, we briefly summarize the method presented in Ravve and Koren (2019) for computing the spatial and directional gradients and Hessians of the ray velocity, which are needed to compute the corresponding derivatives of the traveltime. The gradients and Hessians of the ray velocity are needed to compute the corresponding gradients and Hessians of the Lagrangian provided in Appendix F.

## THE ARCLENGTH-RELATED LAGRANGIAN

Consider a 3D smooth heterogeneous general anisotropic medium and a given initial-guess "ray" trajectory (non-stationary approximated ray path) between two fixed endpoints, $S$ and $R$. Let $\mathbf{x}(s) = \begin{bmatrix} x_1 & x_2 & x_3 \end{bmatrix}$ be a point along the approximated ray path, and $\dot{\mathbf{x}}(s) \equiv d\mathbf{x}(s)/ds \equiv \mathbf{r}(s) = \begin{bmatrix} r_1 & r_2 & r_3 \end{bmatrix}$ the direction vector at this point, normalized to the unit length, $\mathbf{r} \cdot \mathbf{r} = 1$, where $ds$ is an infinitesimal elementary arclength along the path. Note that $\mathbf{r}$ is also the direction of the ray velocity vector. Fermat's principle for a stationary traveltime path can be stated with the use of the Lagrangian $L(\mathbf{x}, \mathbf{r})$ that depends on the position $\mathbf{x}(s)$ and direction $\mathbf{r}(s)$ along the path,

$$t = \int_S^R L(\mathbf{x}, \mathbf{r})\, ds \; \rightarrow \text{stationary} \qquad . \tag{1}$$

In this study, we propose the following arclength-related Lagrangian,

$$L(\mathbf{x}, \mathbf{r}) = \frac{dt}{ds} = \frac{\sqrt{\mathbf{r} \cdot \mathbf{r}}}{v_{\text{ray}}(\mathbf{x}, \mathbf{r})} = \frac{\sqrt{\mathbf{r} \cdot \mathbf{r}}}{v_{\text{ray}}\left[\mathbf{C}(\mathbf{x}), \mathbf{r}\right]} \qquad . \tag{2}$$

The proposed Lagrangian is a first-degree homogeneous function wrt the ray direction $\mathbf{r}$ (see Part II where we provide a comprehensive discussion about the homogeneity degree of the proposed Lagrangian and alternative ones). Among other alternative Lagrangians that we analyzed, we consider this one the most convenient and efficient for our finite element implementation of the Eigenray kinematic (Part III) and dynamic (Part VI) methods in smooth heterogeneous general isotropic and anisotropic media.

The value of the square root $\sqrt{\mathbf{r}\cdot\mathbf{r}}$ in the numerator of the Lagrangian is 1, but we do not replace it by the constant value because it affects the partial derivatives needed for the Euler-Lagrange formulation. The ray velocity magnitude in the denominator, $v_{\text{ray}}\left[\mathbf{C}(\mathbf{x}),\mathbf{r}\right]$, depends implicitly on the location components $\mathbf{x}$, where $\mathbf{C}(\mathbf{x})$ is the medium density-normalized elasticity matrix, and explicitly on the ray direction components (the latter is in particular important in anisotropic media). Note, that the ray velocity magnitude is a zero-degree homogeneous function wrt the tangent vector $k\,\mathbf{r}$, which means that it only depends of the normalized direction vector $\mathbf{r}$ but not on its length $k$.

In Part II of this study we prove mathematically the correctness of the proposed Lagrangian (equation 2) and validate it with different anisotropic scenarios. We also show in Part II that choosing the Lagrangian of equation 2, with $\sqrt{\mathbf{r}\cdot\mathbf{r}}$ in the numerator, imposes using a normalized directional derivative of the ray velocity which is related to the non-normalized one via a special transformation tensor, $\mathbf{T}=\mathbf{I}-\mathbf{r}\otimes\mathbf{r}$ (Ravve and Koren, 2019). This transformation tensor (operator) plays an important role in this study, in both kinematic and dynamic problems.

<u>Remark 1:</u> An alternative Lagrangian (e.g., Červený 2002a, 2002b), with a unity in the numerator (instead of $\sqrt{\mathbf{r}\cdot\mathbf{r}}$), does not require such a normalization, but it leads to virtual, non-physical dependencies of the ray velocity of isotropic and anisotropic media on the length $k$ of the tangent vector $k\mathbf{r}$ specifying the ray velocity direction. In addition, it also leads to a virtual dependence of the isotropic velocity on the ray direction vector $\mathbf{r}$ which is an artificial (non-physical) dependency as well. A detailed discussion is given in Part II. For these reasons, we prefer our proposed form of the Lagrangian.

Remark 2: As mentioned, the form of the Lagrangian is not unique. Moreover, The Lagrangian is usually defined for the traveltime as the flow (characteristic) parameter, rather than the arclength (e.g., Červený 2000, 2002a and 2002b; Slawinski 2015). In Part II we elaborate on the alternative Lagrangians and their relation to the proposed one.

Remark 3: Note that in a regular (non-parametric) functional, e.g., $\int_{x_1}^{x_2} f\left[y(x), y'(x), x\right] dx$, the endpoint values of the argument, $x_1$ and $x_2$, are fixed. This is not so for the parametric functional in equation 1, where the source and receiver spatial locations, $S$ and $R$, are fixed, but the full arclength of the ray trajectory is unknown until the stationary path is found.

Notations for the derivatives wrt the flow parameter:

A parameter with an upper dot and with no subscript means a derivative of the parameter wrt the arclength $s$ (in the dynamic analysis – wrt the arclength of the central ray). A parameter with an upper dot and with a subscript means a derivative wrt the flow parameter indicated by that subscript. In particular, $\dot{\mathbf{x}} \equiv \dot{\mathbf{x}}_s$, $\dot{\mathbf{x}}_\tau$ and $\dot{\mathbf{x}}_\zeta$ represent the derivatives of the ray path location wrt the arclength $s$, current time $\tau$ and a generic flow parameter $\zeta$, respectively. In the finite-element implementation (Parts III and VI), we also use the symbol "prime" (instead of the upper dot) for derivatives wrt the internal flow parameter $-1 \le \xi \le +1$ within any individual element (between the nodes). In particular, $\mathbf{x}'$ and $\mathbf{r}'$ are derivatives of the location and direction wrt the internal parameter $\xi$.

Following equations 1 and 2, there are two ways to derive the kinematic variational formulation, that yield the same first-order algebraic equation set, where the solution (the discretized spatial coordinates and directions of the path) represents the optimized stationary ray . The first approach involves directly the vanishing traveltime gradient computed from the Lagrangian (equation 2) and is described (derived) in Part III. The second approach, which is formulated in this part and further developed in Part III, involves obtaining first the nonlinear, second-order Euler-Lagrange equation, using the proposed Lagrangian, and then applying the weak formulation with the Galerkin method. We explicitly show in Part III that when using the same interpolation scheme for the values between the nodes (in our case, the Hermite polynomial interpolation), the algebraic equations obtained from the two approaches are identical.

## EULER-LAGRANGE EQAUTION

In this section we obtain the main result of this study: the nonlinear second-order Euler-Lagrange kinematic ray equation which is solved in Part III with the finite-element approach.

For the gradients of the Lagrangian $L(\mathbf{x},\mathbf{r})$, we apply the shorthand notations ,

$$L_{\mathbf{x}} = \nabla_{\mathbf{x}} L = \frac{\partial L}{\partial \mathbf{x}} \quad \text{and} \quad L_{\mathbf{r}} = \nabla_{\mathbf{r}} L = \frac{\partial L}{\partial \mathbf{r}} \quad . \tag{3}$$

With these notations, the Euler-Lagrange equation reads,

$$\frac{d}{ds} L_{\mathbf{r}} = L_{\mathbf{x}} \quad . \tag{4}$$

The partial derivatives of $L(\mathbf{x},\mathbf{r})$ are given by,

$$L_{\mathbf{x}} = -\frac{\nabla_{\mathbf{x}} v_{\text{ray}}}{v_{\text{ray}}^2}\sqrt{\mathbf{r}\cdot\mathbf{r}} \quad \text{and} \quad L_{\mathbf{r}} = \frac{\mathbf{r}}{v_{\text{ray}}\sqrt{\mathbf{r}\cdot\mathbf{r}}} - \frac{\nabla_{\mathbf{r}} v_{\text{ray}}}{v_{\text{ray}}^2}\sqrt{\mathbf{r}\cdot\mathbf{r}} \quad . \tag{5}$$

Introduction of equation 5 into equation 4 leads to the explicit form of the nonlinear, second-order, vector-form, Euler-Lagrange ordinary differential equation (ODE),

$$\frac{d}{ds}\left(\frac{\mathbf{r}}{v_{\text{ray}}\sqrt{\mathbf{r}\cdot\mathbf{r}}} - \frac{\nabla_{\mathbf{r}} v_{\text{ray}}}{v_{\text{ray}}^2}\sqrt{\mathbf{r}\cdot\mathbf{r}}\right) = -\frac{\nabla_{\mathbf{x}} v_{\text{ray}}}{v_{\text{ray}}^2}\sqrt{\mathbf{r}\cdot\mathbf{r}} \quad . \tag{6}$$

We note that whenever there is no need to further differentiate wrt the ray direction $\mathbf{r}$, equation 5 can be simplified to,

$$L_{\mathbf{x}} = -\frac{\nabla_{\mathbf{x}} v_{\text{ray}}}{v_{\text{ray}}^2} \quad \text{and} \quad L_{\mathbf{r}} = \frac{\mathbf{r}}{v_{\text{ray}}} - \frac{\nabla_{\mathbf{r}} v_{\text{ray}}}{v_{\text{ray}}^2} \quad , \tag{7}$$

and the actual nonlinear, second-order, Euler-Lagrange ODE to be solved is given by,

$$\frac{d}{ds}\left(\frac{\mathbf{r}}{v_{\text{ray}}} - \frac{\nabla_{\mathbf{r}} v_{\text{ray}}}{v_{\text{ray}}^2}\right) = -\frac{\nabla_{\mathbf{x}} v_{\text{ray}}}{v_{\text{ray}}^2} \quad , \tag{8}$$

which is the principle result of this paper. The expression in the brackets, $L_{\mathbf{r}}$, is the slowness vector, $\mathbf{p}$, consisting of the tangent (to the ray) and normal components (see the proof and details in Appendix A). Relationship $L_{\mathbf{r}} = \mathbf{p}$ constitutes the momentum equation. Part III is dedicated to the solution of equation 8 using the finite element approach, applying the weak formulation and the weighted-residual method (e.g., Galerkin, 1915; Zienkiewicz et al., 2013).

In Appendix B we open the brackets on the left-hand side of equation 6 in order to explicitly obtain all its components in terms of the ray velocity and its derivatives, including the second derivatives. This operation, however, is not needed for the kinematic Eigenray solution; instead, we apply the weak formulation to equation 8 to be solved with the finite element method. The weak formulation effectively eliminates the second derivative of the position vector (or the first derivative of the ray direction), $\ddot{\mathbf{x}} = \dot{\mathbf{r}}$.

## THE ARCLENGTH-RELATED HAMILTONIAN

In this section we provide the arclength-related Hamiltonian $H(s)$, that matches the proposed Lagrangian $L(s)$ through the Legendre transform. The detailed derivation is given in Appendix C.

Consider a general Hamiltonian $H^{\zeta}(\mathbf{x},\mathbf{p})$, where the superscript $\zeta$ indicates an arbitrary flow parameter and its value and units depend on the form of the Hamiltonian. With the shorthand notations for the Hamiltonian derivatives (gradients) wrt the position and slowness components,

$$\frac{\partial H^{\zeta}}{\partial \mathbf{x}} = \nabla_{\mathbf{x}} H^{\zeta} = H^{\zeta}_{\mathbf{x}} \quad , \quad \frac{\partial H^{\zeta}}{\partial \mathbf{p}} = \nabla_{\mathbf{p}} H^{\zeta} = H^{\zeta}_{\mathbf{p}} \quad , \tag{9}$$

the kinematic ray tracing equations can be written as,

$$\frac{d\mathbf{x}}{d\zeta} = H^{\zeta}_{\mathbf{p}} \quad , \quad \frac{d\mathbf{p}}{d\zeta} = -H^{\zeta}_{\mathbf{x}} \quad . \tag{10}$$

We consider the Christoffel equation, valid for general anisotropic media, as a *reference* (vanishing, unitless) ray tracing Hamiltonian,

$$H^{\bar{\tau}} = \det\left[\mathbf{\Gamma} - \mathbf{I}\right] \quad , \qquad \mathbf{\Gamma} = \mathbf{p}\cdot\tilde{\mathbf{C}}\cdot\mathbf{p} \quad , \qquad H^{\bar{\tau}}\left(\mathbf{x},\mathbf{p}\right) = 0 \qquad , \tag{11}$$

where $\mathbf{\Gamma}$ is the Christoffel matrix, $\mathbf{I}$ is the $3\times3$ identity matrix, and $\tilde{\mathbf{C}}$ is the density-normalized fourth-order stiffness (elastic) tensor. The tilde above $\mathbf{C}$ is used to distinguish between the fourth-order tensor $\tilde{\mathbf{C}}$ from its matrix representation, $\mathbf{C}$. The bar above the superscript index $\tau$, indicates that the flow parameter of the Hamiltonian is a "scaled time" $\bar{\tau}$ (rather than the actual time $\tau$), and it has the units of time $\left[\mathrm{T}\right]$. In Appendix C we relate it to the actual traveltime $\tau$, using a unitless scaler, $\alpha\left(\bar{\tau}\right)$. We also show in Appendix C that for the flow parameter arclength, $\zeta \to s$, the ray tracing equation set 10 can be written as,

$$\frac{d\mathbf{x}}{ds} = \mathbf{r} = \frac{H^{\bar{\tau}}_{\mathbf{p}}\left(\mathbf{x},\mathbf{p}\right)}{\sqrt{H^{\bar{\tau}}_{\mathbf{p}}\cdot H^{\bar{\tau}}_{\mathbf{p}}}} \quad , \qquad \frac{d\mathbf{p}}{ds} = -\frac{H^{\bar{\tau}}_{\mathbf{x}}\left(\mathbf{x},\mathbf{p}\right)}{\sqrt{H^{\bar{\tau}}_{\mathbf{p}}\cdot H^{\bar{\tau}}_{\mathbf{p}}}} \qquad . \tag{12}$$

We then introduce the following arclength-related Hamiltonian, connected to the reference Hamiltonian,

$$H\left(\mathbf{x},\mathbf{p}\right) \equiv H^{s}\left(\mathbf{x},\mathbf{p}\right) = \frac{H^{\bar{\tau}}\left(\mathbf{x},\mathbf{p}\right)}{\sqrt{H^{\bar{\tau}}_{\mathbf{p}}\cdot H^{\bar{\tau}}_{\mathbf{p}}}} \qquad . \tag{13}$$

Comment: The arclength-related Hamiltonian and Lagrangian, $H$ and $L$, respectively, are related via the Legendre transform, $H\left(\mathbf{x},\mathbf{p}\right) = L_{\mathbf{r}}\left(\mathbf{x},\mathbf{r}\right)\cdot\mathbf{r} - L\left(\mathbf{x},\mathbf{r}\right)$ (see equation 30 of Part II). The arclength-related Hamiltonian, $H\left(\mathbf{x},\mathbf{p}\right)$, vanishes along the ray, which leads to, $L_{\mathbf{r}}(\mathbf{x},\mathbf{r})\cdot\mathbf{r} = L(\mathbf{x},\mathbf{r})$ (or, equivalently, $\mathbf{p}\cdot\mathbf{r} = v_{\text{ray}}^{-1}$). According to Euler theorem, this property is an evidence that the proposed Lagrangian, $L(\mathbf{x},\mathbf{r})$, is a first-degree homogeneous function wrt the ray direction vector $\mathbf{r}$ (see detailed explanations in Part II).

Since both Hamiltonians, $H^{\overline{\tau}}(\mathbf{x},\mathbf{p})$ and $H(\mathbf{x},\mathbf{p})$, vanish along the ray, the gradients of the arclength-related Hamiltonian are,

$$H_{\mathbf{x}} = \frac{H_{\mathbf{x}}^{\overline{\tau}}}{\sqrt{H_{\mathbf{p}}^{\overline{\tau}}\cdot H_{\mathbf{p}}^{\overline{\tau}}}} \quad , \qquad H_{\mathbf{p}} = \frac{H_{\mathbf{p}}^{\overline{\tau}}}{\sqrt{H_{\mathbf{p}}^{\overline{\tau}}\cdot H_{\mathbf{p}}^{\overline{\tau}}}} \quad , \tag{14}$$

and equation set 12 for the kinematic ray tracing simplifies to,

$$\mathbf{r} = \frac{d\mathbf{x}}{ds} = H_{\mathbf{p}} \quad , \quad \frac{d\mathbf{p}}{ds} = -H_{\mathbf{x}} \quad . \tag{15}$$

A summary of the notations and definitions for the Hamiltonians and Lagrangians used in all parts of this study is given in Tables 1 and 2.

## KINEMATIC EQUATIONS IN TERMS OF RAY VELOCITY AND ITS DERIVATIVES

For completeness, in this section we present the Euler-Lagrange second-order ODE (equation 8) as a set of two first-order kinematic ray tracing ODE. Although we don't use these equations in the kinematic Eigenray solution, they are later used to validate the proposed Lagrangian. The derivation in this section is based on the result obtained in Appendix A, where we prove that the expression in the brackets in equation 8 is the slowness vector $\mathbf{p}$,

$$\mathbf{p} = \frac{\mathbf{r}}{v_{\text{ray}}} - \frac{\nabla_{\mathbf{r}} v_{\text{ray}}}{v_{\text{ray}}^2} \quad , \quad \mathbf{p} = L_{\mathbf{r}} \qquad . \tag{16}$$

In other words, the generalized momentum of the proposed traveltime integrand $L(\mathbf{x},\mathbf{r})$ is the slowness vector $\mathbf{p}$. This is an essential argument for the correctness of the proposed traveltime integrand $L$ as the Lagrangian for general anisotropic media. This relation, $L_{\mathbf{r}} = \mathbf{p}$ , leads to a set of two first-order kinematic ray tracing ODE for the stationary ray path in terms of the ray velocity and its spatial and directional gradients,

$$\frac{d\mathbf{x}}{ds} = \mathbf{p}\, v_{\text{ray}} + \frac{\nabla_{\mathbf{r}} v_{\text{ray}}}{v_{\text{ray}}} \quad , \quad \frac{d\mathbf{p}}{ds} = -\frac{\nabla_{\mathbf{x}} v_{\text{ray}}}{v_{\text{ray}}^2} \qquad . \tag{17}$$

Note that this form includes, in addition to the location vector $\mathbf{x}$, two additional mutually dependent vectors: $\mathbf{p} = \mathbf{p}(\mathbf{x},\mathbf{r})$ and $\mathbf{r} = \mathbf{r}(\mathbf{x},\mathbf{p})$, the slowness and the ray velocity gradient vectors. This dependence is explained later in this paper and expressed by equation 18 that can be considered complementary to equation set 17. In this study however, we formulate the governing kinematic and dynamic equations in terms of the ray velocity alone. Rather than solving equation set 17, we start with equation 8, applying the weak formulation that effectively reduces its second-order ODE to a first-order weighted residual algebraic equation set.

<u>Comment:</u> After we proved that the generalized momentum is the slowness vector, $L_{\mathbf{r}} = \mathbf{p}$, the factor $\sqrt{\mathbf{r}\cdot\mathbf{r}} = 1$ can be kept in the numerator of the arclength-related Lagrangian $L(\mathbf{x},\mathbf{r})$ or be removed, as long as there is no further differentiation wrt the ray direction $\mathbf{r}$, which is the case for the remaining (i.e., other than $L_{\mathbf{r}} = \mathbf{p}$) kinematic ray equation . In either case (with or without the factor $\sqrt{\mathbf{r}\cdot\mathbf{r}}$ ) we obtain the second kinematic equation of set 15, $d\mathbf{p}/ds = L_{\mathbf{x}} = -\nabla_{\mathbf{x}} v_{\text{ray}} / v_{\text{ray}}^2$

.

## COMPUTING THE SLOWNESS VECTOR GIVEN THE RAY DIRECTION

The ray direction vector $\mathbf{r}$ and the slowness vector $\mathbf{p}$ are dependent ray characteristics and should match each other. We distinguish between the forward problem of finding $\mathbf{r}(\mathbf{x},\mathbf{p})$ and the inverse problem of finding $\mathbf{p}(\mathbf{x},\mathbf{r})$. The forward problem is simpler, but in the proposed Eigenray approach, the ray location $\mathbf{x}$ and its direction $\mathbf{r}$ are the primary (input) DoF, while the slowness vector $\mathbf{p}$ is a dependent parameter to be established; thus, we mostly deal with this kind of inverse problem.

### The forward problem

The forward problem can be solved directly, applying the gradient of the arclength-related Hamitonian wrt the slowness vector, $H_{\mathbf{p}} = \mathbf{r}$ (this is also one of the KRT equations). The ray direction can be also obtained with the momentum equation, $L_{\mathbf{r}} = \mathbf{p}$ (as we demonstrate in Appendix H of Part II, with a numerical example for a triclinic medium), but this way is accompanied with unnecessary complications; we do not recoomend such an approach. The reason is that the directional gradient of the Lagrangian, $L_{\mathbf{r}}$, depends on the corresponding normalized directional gradient of the ray velocity magnitude, $\nabla_{\mathbf{r}} v_{\text{ray}}$, which, in turn, is a function of both, the slowness vector, $\mathbf{p}$, and the ray velocity vector, $\mathbf{v}_{\text{ray}}$.

### The inverse problem

For a given wave type and a ray velocity direction $\mathbf{r}$, finding the components of the slowness vector $\mathbf{p}$ is a (nontrivial) inverse problem. This is actually the core problem in solving the Lagrangian-based kinematic problem, where the ray position and direction are the primary variables and the slowness is to be found. (Recall that for shear waves, several solutions for $\mathbf{p}(\mathbf{x},\mathbf{r})$ co-exist.) In this study, we apply the Hamiltonian-based approach, exploiting the collinearity of the Hamiltonian gradient (wrt the slowness vector) and the ray direction (e.g., Musgrave, 1954; Fedorov, 1968; Grechka, 2017), along with the condition for the vanishing Hamiltonian,

$$H_{\mathbf{p}}^{\bar{\tau}}(\mathbf{x},\mathbf{p})\times\mathbf{r}=0 \quad , \qquad H^{\bar{\tau}}(\mathbf{x},\mathbf{p})=0 \qquad . \tag{18}$$

In this equation set, the reference Hamiltonian $H^{\bar{\tau}}(\mathbf{x},\mathbf{p})$ (defined in equation 11) can be replaced by any other Hamiltonian (e.g., by the arclength-related Hamiltonian, $H \equiv H^s$); however, we consider the reference Hamiltonian $H^{\bar{\tau}}$ the simplest for this problem. After the slowness vector is found, we compute the ray velocity magnitude,

$$v_{\text{ray}}=\frac{1}{\mathbf{p}\cdot\mathbf{r}} \qquad . \tag{19}$$

In Appendix D, we provide more details on the technique used to solve equation set 18.

## SPATIAL AND DIRECTIONAL DERIVATIVES OF THE RAY VELOCITY

The optimization scheme includes a discretization of the path into a number of nodes, where each node has a position and a ray velocity direction, with an interpolation between the nodes.

For a stationary kinematic solution, the total traveltime gradient, consisting of spatial and directional blocks, vanishes. In Part III we describe the Newton optimization scheme which also requires the global traveltime Hessian matrix that includes spatial, directional and mixed (spatial/directional) blocks. The global traveltime Hessian matrix is further used to analyze the type of the stationary rays (minimum or saddle point traveltime), for the computation of the dynamic properties, such as the geometric spreading. The dynamic solutions are further used for the identification and classification of caustics.

These first and second derivatives of the traveltime wrt the DoF are the core computation components of the proposed method. In a recent work (Koren and Ravve, 2018b), we showed that these traveltime derivatives are directly related to the corresponding derivatives of the ray (group) velocity, and in the study (Ravve and Koren, 2019), we explained in detail their computation. Due to their central importance in this study, we briefly summarize in Appendix E the main results derived in the latter paper.

## CONCLUSIONS

In this part of our study we establish the theoretical background of our proposed variational (kinematic) Eigenray method, based on Fermat's principle, for obtaining stationary ray paths between two fixed endpoints, in general 3D smooth heterogeneous anisotropic media. We propose an original arclength-related Lagrangian, depending on both the location and direction of the ray trajectory which allows efficient finite element implementation. We first prove that the generalized momentum derived for this Lagrangian (its derivative with respect to the ray direction) is the slowness vector. We also provide the corresponding Hamiltonian for the proposed Lagrangian; the two are related by the Legendre transform. The main result of this study is the derived second-order, Lagrangian-based, ordinary differential equation for the kinematic ray tracing. This equation is obtained in a convenient form suitable for the weak variational formulation, which is then solved in Part III using the finite element approach.

## ACKNOWLEDGEMENT

The authors are grateful to Emerson for the financial and technical support of this study and for the permission to publish its results. The gratitude is extended to Ivan Pšenčík, Einar Iversen, Michael Slawinski, Alexey Stovas, Vladimir Grechka, and our colleague Beth Orshalimy, whose valuable remarks helped to improve the content and style of this paper.

## APPENDIX A.

## GENERALIZED MOMENTUM OF THE ARCLENGTH-RELATED LAGRANGIAN

The generalized momentum equation is defined as the derivative, $L_{\dot{\mathbf{x}}_\zeta}$ , of the chosen Lagrangian $L\left(\mathbf{x},\dot{\mathbf{x}}_\zeta\right)$ wrt $\dot{\mathbf{x}}_\zeta = d\mathbf{x}/d\zeta$, where $\zeta$ is an arbitrary flow parameter along the path. In the Eigenray formulation the flow parameter is the arclength, $\zeta = s$, thus, $d\mathbf{x}/d\zeta = d\mathbf{x}/ds = \mathbf{r}\left(s\right)$ is the ray velocity direction vector. Using the proposed Lagrangian (equation 2), the generalized momentum vector reads,

$$L_{\mathbf{r}} = \frac{\mathbf{r}}{v_{\text{ray}}\sqrt{\mathbf{r}\cdot\mathbf{r}}} - \frac{\nabla_{\mathbf{r}} v_{\text{ray}}}{v_{\text{ray}}^2}\sqrt{\mathbf{r}\cdot\mathbf{r}} \quad . \qquad \text{(A1)}$$

As mentioned, in the case where the momentum vector $L_{\mathbf{r}}$ is not further subjected to additional differentiation wrt the ray direction $\mathbf{r}$, we can apply the normalization rule, $\mathbf{r} \cdot \mathbf{r} = 1$, and the generalized momentum in equation A1 simplifies to,

$$L_{\mathbf{r}} = \frac{\mathbf{r}}{v_{\text{ray}}} - \frac{\nabla_{\mathbf{r}} v_{\text{ray}}}{v_{\text{ray}}^2} \quad . \tag{A2}$$

The aim of this appendix is to prove that the expression on the right side of equation A2 is the slowness vector $\mathbf{p}$.

We start the proof by emphasizing that $\nabla_{\mathbf{r}} v_{\text{ray}}$ is the normalized directional gradient of the ray velocity which is not equal to the non-normalized gradient, $\partial v_{\text{ray}} / \partial \mathbf{r}$ (Ravve and Koren, 2019),

$$\frac{\partial}{\partial \mathbf{r}} v_{\text{ray}} = -v_{\text{ray}}^2 \mathbf{p} \quad . \tag{A3}$$

To avoid confusion, we note that the “normalized” directional gradient $\nabla_{\mathbf{r}} v_{\text{ray}}$ does not mean that it has a unit length; it simply means that the corresponding ray velocity direction $\mathbf{r}$ is forced to have a unit length when computing the gradient components of the ray velocity. On the other hand, the non-normalized derivative $\partial v_{\text{ray}} / \partial \mathbf{r}$ means that this is just a set of partial derivatives wrt $r_i$, keeping the other two ray direction components $r_j$, $j \neq i$, fixed. However, even an infinitesimal change of $r_i$ , with the other two components fixed, ruins the normalization, $\mathbf{r} \cdot \mathbf{r} = 1$. To keep the normalization, Ravve and Koren (2019) suggest the following linear transform,

$$\underbrace{\nabla_{\mathbf{r}} v_{\text{ray}}}_{\text{normalized}} = \left(\mathbf{I} - \mathbf{r} \otimes \mathbf{r}\right) \cdot \underbrace{\frac{\partial v_{\text{ray}}}{\partial \mathbf{r}}}_{\text{non-normalized}} \quad , \qquad \text{(A4)}$$

where the expression in the brackets is a $3 \times 3$ transform matrix (tensor). We emphasize that this difference between the normal and non-normal gradient vectors (equation A4) is a fundemental concept in the formulation of the Eigenray method.

Combining equations A3 and A4, we obtain the ("normalized") directional gradient of the ray velocity $\nabla_{\mathbf{r}} v_{\text{ray}}$,

$$\nabla_{\mathbf{r}} v_{\text{ray}} = -v_{\text{ray}}^2 \left(\mathbf{I} - \mathbf{r} \otimes \mathbf{r}\right) \mathbf{p} = -v_{\text{ray}}^2 \mathbf{p} + v_{\text{ray}}^2 \left(\mathbf{r} \otimes \mathbf{r}\right) \mathbf{p} \quad . \qquad \text{(A5)}$$

From now on we refer to the normalized directional gradient as the directional gradient, omitting the word "normalized". Next, we apply an auxiliary algebraic identity: for any three vectors $\mathbf{a}, \mathbf{b}, \mathbf{c}$ ,

$$\left(\mathbf{a} \otimes \mathbf{b}\right) \cdot \mathbf{c} = \left(\mathbf{a} \otimes \mathbf{c}\right) \cdot \mathbf{b} = \left(\mathbf{b} \cdot \mathbf{c}\right) \mathbf{a} \quad . \qquad \text{(A6)}$$

With the use of this rule, the directional gradient becomes,

$$\nabla_{\mathbf{r}} v_{\text{ray}} = -v_{\text{ray}}^2 \mathbf{p} + v_{\text{ray}}^2 \left(\mathbf{r} \cdot \mathbf{p}\right) \mathbf{r} \quad . \qquad \text{(A7)}$$

Recall that for general anisotropic media,

$$\mathbf{v}_{\text{ray}} \cdot \mathbf{p} = 1 \quad \rightarrow \quad \mathbf{r} \cdot \mathbf{p} = v_{\text{ray}}^{-1} \quad . \qquad \text{(A8)}$$

hence, the directional gradient simplifies to,

$$\nabla_{\mathbf{r}} v_{\text{ray}} = v_{\text{ray}}\mathbf{r} - v_{\text{ray}}^2\mathbf{p} = \mathbf{v}_{\text{ray}} - v_{\text{ray}}^2\mathbf{p} \qquad . \tag{A9}$$

It follows from equation A9 that the directional gradient of the ray velocity is normal to the ray direction,

$$\nabla_{\mathbf{r}} v_{\text{ray}} \cdot \mathbf{r} = 0 \qquad . \tag{A10}$$

Applying a general formula for the gradient of the scalr product,

$$\nabla(\mathbf{a}\cdot\mathbf{b}) = (\nabla\mathbf{a})^T\,\mathbf{b} + (\nabla\mathbf{b})^T\,\mathbf{a} \qquad , \tag{A11}$$

where $\mathbf{a}$ and $\mathbf{b}$ are two arbitrary vectros, we obtain important sequences from equation A10,

$$\begin{gathered} \nabla_{\mathbf{r}} v_{\text{ray}} \cdot \mathbf{r} = 0 \quad \rightarrow \qquad \nabla_{\mathbf{r}}\nabla_{\mathbf{x}} v_{\text{ray}} \cdot \mathbf{r} = 0 \ , \\ \nabla_{\mathbf{r}}\nabla_{\mathbf{r}} v_{\text{ray}} \cdot \mathbf{r} + \nabla_{\mathbf{r}} v_{\text{ray}} = 0 \ , \qquad \mathbf{r}\cdot\nabla_{\mathbf{x}}\nabla_{\mathbf{r}} v_{\text{ray}} = 0 \ , \\ \mathbf{r}\cdot\nabla_{\mathbf{r}}\nabla_{\mathbf{r}} v_{\text{ray}} \cdot \mathbf{r} = 0 \ . \end{gathered} \tag{A12}$$

Finally, the introduction of equation A9 into A2 leads to,

$$L_{\mathbf{r}} = \frac{\mathbf{r}}{v_{\text{ray}}} - \frac{1}{v_{\text{ray}}^2}\left(v_{\text{ray}}\mathbf{r} - v_{\text{ray}}^2\mathbf{p}\right) = \mathbf{p} \qquad . \tag{A13}$$

Thus, we proved that the generalized momentum represents the slowness vector, $L_{\mathbf{r}} = \mathbf{p}$, and

$$\mathbf{p} = \frac{\mathbf{r}}{v_{\text{ray}}} - \frac{\nabla_{\mathbf{r}} v_{\text{ray}}}{v_{\text{ray}}^2} \quad \text{or} \quad \mathbf{p} = \frac{\mathbf{v}_{\text{ray}} - \nabla_{\mathbf{r}} v_{\text{ray}}}{v_{\text{ray}}^2} \qquad . \tag{A14}$$

Equation A14 demonstrates that the slowness vector $\mathbf{p}$ consists of two components: one component is along the ray, and we name it “the group slowness vector”, while the other component is the projection of the slowness vector onto the plane normal to the ray, as shown in Figure 1,

$$(\mathbf{p}\cdot\mathbf{r})\mathbf{r}=\frac{\mathbf{r}}{v_{\text{ray}}} \quad , \qquad \mathbf{r}\times\mathbf{p}\times\mathbf{r}=-\frac{\nabla_{\mathbf{r}}v_{\text{ray}}}{v_{\text{ray}}^2} \quad . \tag{A15}$$

Although the triple cross product is not associative, $(\mathbf{a}\times\mathbf{b})\times\mathbf{c}\neq\mathbf{a}\times(\mathbf{b}\times\mathbf{c})$, it becomes associative for $\mathbf{a}=\mathbf{c}$. Hence, we omit the brackets in the second equation of set A15. Equation A15 shows that the proposed variational formulation relates the ray velocity vector, $\mathbf{v}_{\text{ray}}$, to the slowness vector, $\mathbf{p}$, thus providing the, so-called, phase-space information.

The second equation of set A15 can be also arranged as,

$$\nabla_{\mathbf{r}}v_{\text{ray}}=-\mathbf{v}_{\text{ray}}\times\mathbf{p}\times\mathbf{v}_{\text{ray}} \quad . \tag{A16}$$

Thus, to establish the directional gradient of the ray velocity, both, the slowness and the ray velocity vectors, are required. The directional gradient vector, $\nabla_{\mathbf{r}}v_{\text{ray}}$**,** belongs to the plane that includes both vectors, $\mathbf{p}$ and $\mathbf{v}_{\text{ray}}$, and is also normal to $\mathbf{v}_{\text{ray}}$ (see Figure 1). Hence, the following mixed product vanishes,

$$\mathbf{p}\times\mathbf{v}_{\text{ray}}\cdot\nabla_{\mathbf{r}}v_{\text{ray}}=0 \quad . \tag{A17}$$

Note that in isotropic media, the slowness and the ray velocity vectors are collinear, and this plane does not exist, and thus the directional gradient of the ray velocity does not exist either.

Finally, the Euler-Lagrange equation (equation 8) can now be arranged in the following compact form,

$$\frac{d\mathbf{p}}{ds} = L_{\mathbf{x}} \quad \text{or} \quad \dot{\mathbf{p}} = L_{\mathbf{x}} \qquad . \tag{A18}$$

## APPENDIX B. DIFFERENTIAL EQUATION OF THE STATIONARY RAY PATH

In this appendix, we expand the Euler-Lagrange kinematic ray tracing ODE that results from the stationary traveltime, ,

$$\frac{d}{ds} L_{\mathbf{r}} = L_{\mathbf{x}} \qquad . \tag{B1}$$

We apply the chain rule for the arclength derivative of the slowness vector, $L_{\mathbf{r}} = \mathbf{p}$, where $L_{\mathbf{r}}(s) = L_{\mathbf{r}}\left[\mathbf{x}(s), \mathbf{r}(s)\right]$ depends on the location and direction of the points along the ray,

$$\frac{dL_{\mathbf{r}}}{ds} = \frac{\partial L_{\mathbf{r}}}{\partial \mathbf{x}} \frac{d\mathbf{x}}{ds} + \frac{\partial L_{\mathbf{r}}}{\partial \mathbf{r}} \frac{d\mathbf{r}}{ds} = L_{\mathbf{rx}}\mathbf{r} + L_{\mathbf{rr}}\dot{\mathbf{r}} \qquad . \tag{B2}$$

Combining equations B1, and B2, we obtain the second-order ODE that does not include the slowness vecror,

$$L_{\mathbf{rr}}(\mathbf{x}, \mathbf{r})\dot{\mathbf{r}} = L_{\mathbf{x}}(\mathbf{x}, \mathbf{r}) - L_{\mathbf{rx}}(\mathbf{x}, \mathbf{r})\mathbf{r} \qquad . \tag{B3}$$

The explicit forms of the gradients and the Hessians of the Lagrangian are listed in equation F2. Note that equation B3, in its present form, is not resolvable for the the curvature vector $\dot{\mathbf{r}}$ due to the singularity of the directional Hessian of the Lagrangian, $L_{\mathbf{rr}}$. In other words, the three scalar equations of set B3 are dependent. However, the scalar product $\mathbf{r}\cdot\dot{\mathbf{r}}$ vanishes, because vector $\mathbf{r}$ has a constant unit length, $\mathbf{r}\cdot\mathbf{r}=1$ and can only change its direction (thus, $\mathbf{r}$ and $\dot{\mathbf{r}}$ are normal to each other). With this additional constraint, taking into account that $L_{\mathbf{rr}}$ is a symmetric matrix, set B4 can be arranged as,

$$\left(L_{\mathbf{rr}}^2 + w^2\mathbf{r}\otimes\mathbf{r}\right)\dot{\mathbf{r}} = L_{\mathbf{rr}}\left(L_{\mathbf{x}} - L_{\mathbf{rx}}\mathbf{r}\right) \quad , \tag{B4}$$

where $w$ is an arbitrary positive weight with the units of slowness (its value does not affect the solution). Vector $\left(\mathbf{r}\otimes\mathbf{r}\right)\dot{\mathbf{r}}=\left(\mathbf{r}\cdot\dot{\mathbf{r}}\right)\mathbf{r}$ vanishes due to the above constraint. However, the matrix in the brackets in equation B4 is not singular; this allows experessing the curvature components explicitly, $\dot{\mathbf{r}}=\mathbf{f}\left(\mathbf{x},\mathbf{r}\right)$ or $\ddot{\mathbf{x}}=\mathbf{f}\left(\mathbf{x},\dot{\mathbf{x}}\right)$, and solving the initial-value kinematic problem with a numerical integration technique, such as Runge-Kutta, for the Lagrangian (rather than the Hamiltonian) formulation.

Consider anistropic medium as a particular case. The gradients and Hessians of the Lagrangian for this case are listed in equation F3. Combining this equation with B3, we obtain,

$$\dot{\mathbf{r}} = \frac{\mathbf{r}\otimes\nabla v}{v}\mathbf{r} - \frac{\nabla v}{v} = -\frac{\nabla v - \left(\nabla v\cdot\mathbf{r}\right)\mathbf{r}}{v} \quad . \tag{B5}$$

The numerator on the right-hand side is the difference between the velocity gradient $\nabla v$ and its tangent projection on the ray direction. It represents the projection of the velocity gradient on the plane normal to the ray,

$$\underbrace{\left(\nabla v\cdot\mathbf{r}\right)\mathbf{r}}_{\text{tangent projection}} + \underbrace{\mathbf{r}\times\nabla v\times\mathbf{r}}_{\text{normal projection}} = \underbrace{\nabla v}_{\text{full gradient}} \quad . \tag{B6}$$

The kinematic equation for isotropic media becomes,

$$\dot{\mathbf{r}} = -\frac{\mathbf{r}\times\nabla v\times\mathbf{r}}{v(\mathbf{x})} \quad \text{or} \quad \ddot{\mathbf{x}} = -\frac{\dot{\mathbf{x}}\times\nabla v\times\dot{\mathbf{x}}}{v(\mathbf{x})} \quad , \tag{B7}$$

which is equivalent to the canonical ray tracing equation set. Equation B7 demonstrates that in isotropic media, the curvature of the ray is proportional to the normal counterpart of the velocity gradient.

We re-emphasize that in this work we do not solve the second-order ray tracing equation B4 directly. Rather, we search for a stationary solution of the traveltime integral by applying the weak finite-element formulation to equation 8. Furthermore, the ray tracing equations in the anisotropic media are normally formulated (as initial-value problems) in terms of the slowness-dependent Hamiltonian rather than the ray-velocity-dependent Lagrangian. The second-order ODE B4, written in terms of the trajectory location and direction vectors, is equivalent to two first-order equations in terms of the trajectory location and slowness vectors. Equation B4 has been derived in this appendix only for demonstrating the expanded (explicit) components of equation B1, the Euler-Lagrange second-order, general anisotropic, kinematic ODE.

## APPENDIX C.

## RAY TRACING EQUATIONS WITH THE ARCLENGTH-RELATED HAMILTONIAN

The general ray tracing equations are given in equation set 10,

$$\frac{d\mathbf{x}}{d\zeta} = H^{\zeta}_{\mathbf{p}} \quad , \quad \frac{d\mathbf{p}}{d\zeta} = -H^{\zeta}_{\mathbf{x}} \quad , \tag{C1}$$

where the value and units of the flow parameter $\zeta$ depend on the form of the Hamiltonian, and the shorthand notations are used for the Hamiltonian gradients wrt the position and slowness. For the reference Hamiltonian of equation 11, $H^{\bar{\tau}} = \det\left[\mathbf{\Gamma} - \mathbf{I}\right]$, the flow parameter is the scaled-time $\bar{\tau}$, where its differential can be written as $d\bar{\tau} = \alpha_{\mathrm{sc}}\left(\tau\right) d\tau$; $\tau$ is the actual time and $\alpha_{\mathrm{sc}}\left(\tau\right)$ is a varying unitless scaler. We rearrange equation C1 for the reference Hamiltonian, $\zeta \to \bar{\tau}$ and $H^{\zeta} \to H^{\bar{\tau}}$,

$$\frac{d\mathbf{x}}{d\bar{\tau}} = H_{\mathbf{p}}^{\bar{\tau}} \quad , \quad \frac{d\mathbf{p}}{d\bar{\tau}} = -H_{\mathbf{x}}^{\bar{\tau}} \quad , \qquad \text{(C2)}$$

Using equation C2, we set the relation between $d\bar{\tau}$ and the arclength differential $ds$,

$$\frac{ds}{d\bar{\tau}} = \sqrt{\frac{d\mathbf{x}}{d\bar{\tau}} \cdot \frac{d\mathbf{x}}{d\bar{\tau}}} = \sqrt{H_{\mathbf{p}}^{\bar{\tau}} \cdot H_{\mathbf{p}}^{\bar{\tau}}} = \left|H_{\mathbf{p}}^{\bar{\tau}}\right| \quad . \qquad \text{(C3)}$$

This makes it possible to convert the ray tracing equation from the flow parameter $\bar{\tau}$ to the arclength $s$,

$$\frac{d\mathbf{x}}{ds} = \frac{d\mathbf{x}/d\bar{\tau}}{ds/d\bar{\tau}} = \frac{H_{\mathbf{p}}^{\bar{\tau}}}{\left|H_{\mathbf{p}}^{\bar{\tau}}\right|} \quad , \quad \frac{d\mathbf{p}}{ds} = \frac{d\mathbf{p}/d\bar{\tau}}{ds/d\bar{\tau}} = -\frac{H_{\mathbf{x}}^{\bar{\tau}}}{\left|H_{\mathbf{p}}^{\bar{\tau}}\right|} \quad , \qquad \text{(C4)}$$

The relation between the actual traveltime increment, $d\tau$, and the arclength differential, $ds$, is then given by,

$$\frac{d\tau}{ds} = \frac{d\tau}{d\mathbf{x}} \cdot \frac{d\mathbf{x}}{ds} = \frac{\mathbf{p} \cdot H_{\mathbf{p}}^{\bar{\tau}}}{\left|H_{\mathbf{p}}^{\bar{\tau}}\right|} = \frac{1}{v_{\mathrm{ray}}} \quad , \quad \mathbf{v}_{\mathrm{ray}} = \frac{H_{\mathbf{p}}^{\bar{\tau}}}{\mathbf{p} \cdot H_{\mathbf{p}}^{\bar{\tau}}} \quad . \qquad \text{(C5)}$$

With the use of notation $H \equiv H^S$ and equation 14, equations C4 and C5 can be arranged as,

$$\frac{d\mathbf{p}}{ds} = -H_{\mathbf{x}} \quad , \quad \mathbf{r} = \frac{d\mathbf{x}}{ds} = H_{\mathbf{p}} \quad , \quad \frac{d\tau}{ds} = \mathbf{p} \cdot H_{\mathbf{p}} = \frac{1}{v_{\mathrm{ray}}} \quad . \tag{C6}$$

Equation set C6 (along with the definitions in equation 13) represents the Hamiltonian kinematic ray tracing equations for the flow parameter arclength. Equation C5 for the ray velocity makes it possible also to rearrange the ray tracing equation C2 in terms of the actual traveltime,

$$\frac{d\mathbf{x}}{d\tau} = \mathbf{v}_{\mathrm{ray}} = \frac{d\mathbf{x}/ds}{d\tau/ds} = \frac{H_{\mathbf{p}}^{\bar{\tau}}}{\mathbf{p} \cdot H_{\mathbf{p}}^{\bar{\tau}}} \quad , \quad \frac{d\mathbf{p}}{d\tau} = \frac{d\mathbf{p}/ds}{d\tau/ds} = -\frac{H_{\mathbf{x}}^{\bar{\tau}}}{\mathbf{p} \cdot H_{\mathbf{p}}^{\bar{\tau}}} \quad . \tag{C7}$$

Hence, the scale factor $\alpha_{\mathrm{sc}}$ of the flow parameter $\bar{\tau}$ becomes,

$$\alpha_{\mathrm{sc}}(\tau) = \frac{d\bar{\tau}}{d\tau} = \frac{d\bar{\tau}/ds}{d\tau/ds} = \frac{1}{\mathbf{p} \cdot H_{\mathbf{p}}^{\bar{\tau}}} \quad . \tag{C8}$$

Thus, it becomes suitable to introduce the time-related Hamiltonian $H^{\tau}$,

$$H^{\tau} = H^{\bar{\tau}} \alpha_{\mathrm{sc}}(\tau) = \frac{H^{\bar{\tau}}}{\mathbf{p} \cdot H_{\mathbf{p}}^{\bar{\tau}}} \quad , \tag{C9}$$

and the kinematic ray tracing equation set C7, with the flow parameter (actual) traveltime $\tau$, simplifies to,

$$\frac{d\mathbf{x}}{d\tau} = H_{\mathbf{p}}^{\tau} = \mathbf{v}_{\mathrm{ray}} \quad , \quad \frac{d\mathbf{p}}{d\tau} = -H_{\mathbf{x}}^{\tau} \quad , \quad \frac{ds}{d\tau} = v_{\mathrm{ray}} = \left| H_{\mathbf{p}}^{\tau} \right| \quad . \tag{C10}$$

In a similar manner the scaled sigma-related Hamiltonian $H^{\sigma}$ (see Table 1) and the corresponding kinematic equations can be constructed.

Remark: The Christoffel equation 11 for the vanishing reference Hamiltonian relates the three slowness components: they are not independent, but depend on each other and the medium properties. Normally, the slowness direction is given and its magnitude should be computed, or two slowness components are given and the third should be computed. The Christoffel equation makes it possible to establish the polarization vector $\mathbf{g}$ because the product $(\mathbf{\Gamma}-\mathbf{I})\mathbf{g}$ vanishes. This means that the three lines of the matrix $\mathbf{\Gamma}-\mathbf{I}$ are dependent and coplanar: they all belong to a plane normal to the polarization vector. Apart from some special "accidental" cases, any two rows of this matrix are independent, and thus, their cross product, normalized to the unit length, represents the polarization vector. Note that in the case of shear-wave singularity, the three rows of the matrix $\mathbf{\Gamma}-\mathbf{I}$ are collinear rather than coplanar, i.e., they are all dependent on each other.

## APPENDIX D. COMPUTING THE RAY VELOCITY MAGNITUDE

At each iteration of the Newton method of the Eigenray procedure, the locations $\mathbf{x}$ and directions $\mathbf{r}$ of a trial ray path are given and need to be refined (updated). This iterative step requires the computation of the magnitude of the ray velocity and its spatial and directional derivatives. which in turn, requires first to establish the slowness components.

Given the medium elastic properties at a given location $\mathbf{C}(\mathbf{x})$, and the ray velocity direction $\mathbf{r}$, we compute the corresponding slowness vector by solving the Hamiltonian-based nonlinear set of three polynomial equations (e.g., Musgrave, 1954; Fedorov, 1968), for a given wave mode (equation set 18). For compressional waves, the magnitude of the ray velocity is uniquely defined by its direction, while for shear waves up to 18 solutions may co-exist (Grechka, 2017). We find the reference Hamiltonian $H^{\bar{\tau}}$ (equation 11), the most suitable for this operation (although any other Hamiltonian can be used). Using the fact that the gradient of the Hamiltonian wrt the slowness components, $H_{\mathbf{p}}^{\bar{\tau}}$, is parallel to the ray direction $\mathbf{r}$, and that the Hamiltonian vanishes along the ray, the slowness components can be obtained from the following set,

$$\underbrace{H_{\mathbf{p}}^{\bar{\tau}}\left(\mathbf{x},\mathbf{p}\right)\times\mathbf{r}=0}_{\text{use 2 equations from 3}} \quad , \quad \underbrace{H^{\bar{\tau}}\left(\mathbf{x},\mathbf{p}\right)=0}_{\text{the third equation}} \quad , \tag{D1}$$

Note that for the arclength-related Hamiltonian, the collinearity equation, $H_{\mathbf{p}}^{\bar{\tau}}\left(\mathbf{x},\mathbf{p}\right)\times\mathbf{r}=0$, reduces to, $H_{\mathbf{p}}\left(\mathbf{x},\mathbf{p}\right)=\mathbf{r}$, still, it is more convenient to apply the reference Hamiltonian in set D2. The cross-product in this equation set represents three scalar equations, but only two of them are independent. We therefore discard one of the cross-product components (for example, the third component, provided $r_3 \neq 0$ ), and apply the vanishing Hamiltonian instead (Grechka, 2017), as indicated under the components of set D1.

The reference Hamiltonian $H^{\bar{\tau}}$, in turn, is presented for any medium by,

$$H^{\bar{\tau}}=\det\left[\mathbf{\Gamma}\left(\mathbf{x},\mathbf{p}\right)-\mathbf{I}\right] \, , \quad \mathbf{\Gamma}\left(\mathbf{x},\mathbf{p}\right)=\mathbf{p}\tilde{\mathbf{C}}\left(\mathbf{x}\right)\mathbf{p} \quad . \tag{D2}$$

$\mathbf{\Gamma}$ is the Christoffel matrix (second-order tensor), and $\tilde{\mathbf{C}}\left(\mathbf{x}\right)$ is the density-normalized fourth-order stiffness tensor, whose components are location-dependent.

Set D1 consists initially of four equations: it includes three components of the cross-product, and the vanishing Hamiltonian. Since the cross-product components are dependent, we can solve all four nonlinear equations with the least-squares approach, and the solution will be exact, since the system is not over-defined. This method effectively converts the set of four original equations into a set of three derived equations, without discarding any of them. The advantage is obvious: no need to decide which equation should be discarded. Should we keep all equations, set D1 generates the target function, $f(\mathbf{p})$,

$$f(\mathbf{p}) = \frac{1}{2}\left[H_{\mathbf{p}}^{\bar{\tau}}(\mathbf{x},\mathbf{p}) \times \mathbf{r}\right] \cdot \left[H_{\mathbf{p}}^{\bar{\tau}}(\mathbf{x},\mathbf{p}) \times \mathbf{r}\right] + \frac{w}{2}\left[H^{\bar{\tau}}(\mathbf{x},\mathbf{p})\right]^2 \to \min \quad , \tag{D3}$$

and this minimum is zero, because the four equations are compatible. The weight $w$ is needed only to adjust the units of the two items in the target function; it does not affect the solution. The weight has the units of velocity squared. At the minimum point, the gradient $f_{\mathbf{p}}$ of the target function in equation D3, wrt the slowness vector components, vanishes,

$$f_{\mathbf{p}} = \underbrace{\left[\mathbf{r} \times H_{\mathbf{p}}^{\bar{\tau}}(\mathbf{x},\mathbf{p})\right]}_{\substack{\text{vector, 3} \\ \text{normal to } \mathbf{r}}} \cdot \underbrace{\left[\mathbf{r} \times H_{\mathbf{pp}}^{\bar{\tau}}(\mathbf{x},\mathbf{p})\right]}_{\substack{\text{matrix, } 3\times 3 \\ \text{each column} \\ \text{normal to } \mathbf{r}}} + w \underbrace{H^{\bar{\tau}}(\mathbf{x},\mathbf{p})}_{\text{scalar}} \underbrace{H_{\mathbf{p}}^{\bar{\tau}}(\mathbf{x},\mathbf{p})}_{\text{vector, 3}} = 0 \quad . \tag{D4}$$

Note that relationship D4 includes cross product of a vector and a second-order tensor that may be generally presented as (e.g., Naumenko and Altenbach, 2007) (Section A.4.7),

$$\mathbf{B} = \mathbf{c} \times \mathbf{A} \qquad \text{or} \qquad \mathbf{B}' = \mathbf{A} \times \mathbf{c} \quad , \tag{D5}$$

where $\mathbf{A}, \mathbf{B}$ and $\mathbf{B}'$ are the second-order tensors. In the first case, the columns of the resulting tensor $\mathbf{B}$ are the cross-products of vector $\mathbf{c}$ and the corresponding columns of tensor $\mathbf{A}$. In the second case, the rows of the resulting tensor $\mathbf{B}'$ are the cross-products of the corresponding rows of tensor $\mathbf{A}$ and vector $\mathbf{c}$. The two forms are related to each other, $\mathbf{c}\times\mathbf{A} = -\left(\mathbf{A}^T \times \mathbf{c}\right)^T$. (In a similar manner, the cross product of two second-order tensors can be defined, $\mathbf{U} = \mathbf{A}\times\mathbf{B}$. The resulting third-order tensor $\mathbf{U}$ can be viewed as a matrix, where each cell contains a vector following from a regular vector cross-product of a row from $\mathbf{A}$ and a column from $\mathbf{B}$.) Note that both cross-products, $\mathbf{c}\times\mathbf{A}$ or $\mathbf{A}\times\mathbf{c}$ for arbitrary vector $\mathbf{c}$ and second-order tensor $\mathbf{A}$ result in a singular second-order tensor with a vanishing determinant.

Vector-form set D4 consists of three independent scalar equations. This set is nonlinear and can be solved, for example, by the Newton method, performing the linearization of the gradient, $\nabla_{\mathbf{p}} f \equiv f_{\mathbf{p}}$, at each iteration,

$$f_{\mathbf{pp}}\left(\mathbf{p}_{\mathrm{o}}\right)\Delta\mathbf{p} = -f_{\mathbf{p}}\left(\mathbf{p}_{\mathrm{o}}\right) \quad , \tag{D6}$$

where $\nabla_{\mathbf{p}}\nabla_{\mathbf{p}} f = f_{\mathbf{pp}}$ is the symmetric Hessian matrix of the target function wrt the slowness components, $\mathbf{p}_{\mathrm{o}}$ is the slowness vector obtained at the previous iteration, and $\Delta\mathbf{p}$ is its correction, to be computed from the linearized equation set D6.

The general formula for the gradient of a tensor product, and its particular case when the first factor is a vector, are needed to compute the slowness Hessian,

$$\nabla\left(\mathbf{AB}\right) = \underbrace{\left[\left(\nabla\mathbf{A}\right)^{T\{1,3,2\}}\mathbf{B}\right]^{T\{1,3,2\}} + \mathbf{A}\nabla\mathbf{B}}_{\text{third-order tensor}} \ , \qquad \begin{aligned} \nabla\left(\mathbf{aB}\right) &= \underbrace{\mathbf{B}^T\nabla\mathbf{a} + \mathbf{a}\nabla\mathbf{B}}_{\text{second-order tensor}} \\ \nabla\left(\mathbf{Ba}\right) &= \underbrace{\left(\nabla\mathbf{B}\right)^{T\{1,3,2\}}\mathbf{a} + \mathbf{B}\nabla\mathbf{a}}_{\text{second-order tensor}} \end{aligned} \quad , \tag{D7}$$

where $\mathbf{A}$ and $\mathbf{B}$ are the second-order tensors, and $\mathbf{a}$ is a vector. Recall that the transpose operator $T\{1,3,2\}$ means that the first index of the third-order tensor remainds unchanged, while the two other indices swap.

We will also use the formula for the gradient of the cross product of two vectors,

$$\nabla(\mathbf{a}\times\mathbf{b}) = \mathbf{a}\times\nabla\mathbf{b} - \mathbf{b}\times\nabla\mathbf{a} \quad . \tag{D8}$$

This yields the Hessian matrix of the target function wrt the slowness vector component, $f_{\mathbf{pp}}$,

$$\begin{aligned} f_{\mathbf{pp}} = & \underbrace{\underbrace{\left[\mathbf{r}\times H^{\bar{\tau}}_{\mathbf{pp}}(\mathbf{x},\mathbf{p})\right]^T}_{\text{order 2}} \cdot \underbrace{\left[\mathbf{r}\times H^{\bar{\tau}}_{\mathbf{pp}}(\mathbf{x},\mathbf{p})\right]}_{\text{order 2}}}_{\text{order 2}} + \underbrace{\underbrace{\left[\mathbf{r}\times H^{\bar{\tau}}_{\mathbf{p}}(\mathbf{x},\mathbf{p})\right]}_{\text{vector}} \cdot \underbrace{\left[\mathbf{r}\times H^{\bar{\tau}}_{\mathbf{ppp}}(\mathbf{x},\mathbf{p})\right]}_{\text{order 3}}}_{\text{order 2}} \\ & + w\underbrace{\underbrace{H^{\bar{\tau}}_{\mathbf{p}}(\mathbf{x},\mathbf{p})}_{\text{vector}} \otimes \underbrace{H^{\bar{\tau}}_{\mathbf{p}}(\mathbf{x},\mathbf{p})}_{\text{vector}}}_{\text{order 2}} + w\underbrace{H^{\bar{\tau}}(\mathbf{x},\mathbf{p})}_{\text{scalar}}\underbrace{H^{\bar{\tau}}_{\mathbf{pp}}(\mathbf{x},\mathbf{p})}_{\text{order 2}} \ . \end{aligned} \tag{D9}$$

In Appendix E, we will need, in addition to the pure slowness Hessian, $f_{\mathbf{pp}}$, also two mixed Hessians of the target function, wrt a) the slowness and the ray direction vectors, $\nabla_{\mathbf{p}}\nabla_{\mathbf{r}} f \equiv f_{\mathbf{pr}}$, and b) wrt the slowness and the position vectors, $\nabla_{\mathbf{p}}\nabla_{\mathbf{x}} f \equiv f_{\mathbf{px}}$,

$$f_{\mathbf{pr}} = -\underbrace{\underbrace{\left[\mathbf{r}\times H^{\bar{\tau}}_{\mathbf{pp}}(\mathbf{x},\mathbf{p})\right]^T}_{\text{order 2}} \cdot \underbrace{\left[H^{\bar{\tau}}_{\mathbf{p}}(\mathbf{x},\mathbf{p})\times\mathbf{I}\right]}_{\text{order 2}}}_{\text{order 2}} - \underbrace{\underbrace{\left[\mathbf{r}\times H^{\bar{\tau}}_{\mathbf{p}}(\mathbf{x},\mathbf{p})\right]}_{\text{vector}} \cdot \underbrace{\left[H^{\bar{\tau}}_{\mathbf{pp}}(\mathbf{x},\mathbf{p})\times\mathbf{I}\right]^{T\{2,1,3\}}}_{\text{order 3}}}_{\text{order 2}} \quad , \tag{D10}$$

and,

$$f_{\mathbf{px}} = \underbrace{\underbrace{\left[\mathbf{r}\times H_{\mathbf{pr}}^{\bar{\tau}}(\mathbf{x},\mathbf{p})\right]^T}_{\text{order 2}} \cdot \underbrace{\left[\mathbf{r}\times H_{\mathbf{px}}^{\bar{\tau}}(\mathbf{x},\mathbf{p})\right]}_{\text{order 2}}}_{\text{order 2}} + \underbrace{\underbrace{\left[\mathbf{r}\times H_{\mathbf{p}}^{\bar{\tau}}(\mathbf{x},\mathbf{p})\right]}_{\text{vector}} \cdot \underbrace{\left[\mathbf{r}\times H_{\mathbf{ppx}}^{\bar{\tau}}(\mathbf{x},\mathbf{p})\right]}_{\text{order 3}}}_{\text{order 2}}$$

$$+ w \underbrace{\underbrace{H_{\mathbf{p}}^{\bar{\tau}}(\mathbf{x},\mathbf{p})}_{\text{vector}} \otimes \underbrace{H_{\mathbf{x}}^{\bar{\tau}}(\mathbf{x},\mathbf{p})}_{\text{vector}}}_{\text{order 2}} + w \underbrace{H^{\bar{\tau}}(\mathbf{x},\mathbf{p})}_{\text{scalar}} \underbrace{H_{\mathbf{px}}^{\bar{\tau}}(\mathbf{x},\mathbf{p})}_{\text{order 2}} \quad . \tag{D11}$$

The third-order tensor $\mathbf{r}\times H_{\mathbf{ppp}}^{\bar{\tau}}(\mathbf{x},\mathbf{p})$ can be viewed as a matrix, whose cells include vectors rather than numbers; these vectors result from the cross-products. The first index is related to the components of the cross-product vector, while the two other indices point to the cell in this matrix. The cross product of two matrices (second-order tensors) is the third-order tensor, where the first belong to the lines of the first factor, the third index – to the columns of the second factor, and the second index – to the components of the cross-product vectors.

After the slowness components have been found, we compute the magnitude of the ray velocity,

$$v_{\text{ray}} = \frac{1}{\mathbf{p}\cdot\mathbf{r}} \quad . \tag{D12}$$

To obtain an initial guess $\mathbf{p}_0$ for equation set D4, we start with the (improper) assumtion that the phase and ray directions are identical. A better initial guess (for compressional waves and general anisotropy) can be obtained using the weak-anisotropy approximation for the difference between the ray and phase directions, suggested by Pšenčík and Vavryčuk (2002), improved by Farra (2004), and later applied by Farra and Pšenčík (2013). The authors compute the difference between the ray and phase directions, $\mathbf{r}-\mathbf{n}$, in the local frame of reference, where the phase direction $\mathbf{n}$ is considered vertical, with subsequent rotation of the resulting vector to the global frame. We follow the same computational formulae, but establish the difference vector, $\mathbf{r}-\mathbf{n}$, directly in the global frame. We formulate the governing retionship in the tensor form (which we consider more convenient than the original component-wise form of the cited works),

$$\mathbf{r}-\mathbf{n} \approx 2\frac{\mathbf{n}\times\mathbf{v}\times\mathbf{n}}{\mathbf{n}\cdot\mathbf{v}} \quad , \quad \text{where} \quad \mathbf{v}\equiv\hat{\mathbf{\Gamma}}\mathbf{n} \quad , \quad \hat{\mathbf{\Gamma}}=\mathbf{n}\tilde{\mathbf{C}}\mathbf{n} \quad . \tag{D13}$$

Note that:

- The denominator approximates the phase velocity squared.
- Vector $\mathbf{v}$ appears in a linear form in both, the numerator and denominator; thus, only its direction is essential, and this vector may be normalized to the unit length.
- In the case of isotropic media, the phase direction is also the eigenvector of the Christoffel matrix (the polarization vector) for compressional waves, $\mathbf{g}=\mathbf{n}$, and the double cross product in equation D13 vanishes. For a weak anisotropy, the difference between the ray and phase directions is a short vector, $|\mathbf{n}-\mathbf{r}| \ll 1$, normal to the phase direction $\mathbf{n}$.
- In the Eigenray workflow, we know the ray direction $\mathbf{r}$ and need to approximate the phase direction $\mathbf{n}$. However, for a weak anisotropy, the difference between the two directions in equation D13can be formulated (alternatively) in terms of the ray direction $\mathbf{r}$, and the result will be approximately the same,

$$\mathbf{r}-\mathbf{n} \approx 2\frac{\mathbf{r}\times\mathbf{v_r}\times\mathbf{r}}{\mathbf{r}\cdot\mathbf{v_r}} \quad , \quad \text{where} \quad \mathbf{v_r}=\hat{\mathbf{\Gamma}}_\mathbf{r}\,\mathbf{r} \quad , \quad \hat{\mathbf{\Gamma}}_\mathbf{r}=\mathbf{r}\,\tilde{\mathbf{C}}\,\mathbf{r} \quad , \tag{D14}$$

  and $\hat{\mathbf{\Gamma}}_\mathbf{r}$ approximates the Christoffel matrix.

## APPENDIX E. SPATIAL AND DIRECTIONAL DERIVATIVES OF THE RAY VELOCITY

Recall that after solving equation set D4, we know the slowness vector $\mathbf{p}$, the ray direction $\mathbf{r}$, and the ray velocity magnitude, $v_{\mathrm{ray}}$. In this appendix we further elaborate on the spatial and directional gradients of the ray velocity, $\nabla_{\mathbf{x}} v_{\mathrm{ray}}$, $\nabla_{\mathbf{r}} v_{\mathrm{ray}}$, and the spatial, directional and mixed Hessians $\nabla_{\mathbf{x}}\nabla_{\mathbf{x}} v_{\mathrm{ray}}$, $\nabla_{\mathbf{r}}\nabla_{\mathbf{r}} v_{\mathrm{ray}}$, $\nabla_{\mathbf{x}}\nabla_{\mathbf{r}} v_{\mathrm{ray}}$ and $\nabla_{\mathbf{r}}\nabla_{\mathbf{x}} v_{\mathrm{ray}}$, which are required for the computation of the corresponding spatial and directional gradients and Hessians of the Lagrangian (see Appendix F). Actually, there are only three different ray velocity Hessians, because the two mixed Hessians are transposed to each other,

$$\nabla_{\mathbf{x}}\nabla_{\mathbf{r}} v_{\mathrm{ray}} = \left(\nabla_{\mathbf{r}}\nabla_{\mathbf{x}} v_{\mathrm{ray}}\right)^T \quad . \tag{E1}$$

A recent study by Ravve and Koren (2019) was devoted to the computation of the spatial and directional derivatives of the ray velocity, so we only discuss them here very briefly.

<u>Directional gradient of the ray velocity</u>

An important aspect of the cited paper involves distinguishing between the so-called non-normalized directional derivatives, $\partial v_{\mathrm{ray}} / \partial \mathbf{r}$,

$$\frac{\partial v_{\mathrm{ray}}}{\partial \mathbf{r}} = -v_{\mathrm{ray}}^2 \mathbf{p} \quad , \tag{E2}$$

and the normalized directional gradient, $\nabla_{\mathbf{r}} v_{\mathrm{ray}}$, of the ray velocity,

$$\nabla_{\mathbf{r}} v_{\text{ray}} = \mathbf{T} \frac{\partial v_{\text{ray}}}{\partial \mathbf{r}} \qquad \text{where} \qquad \mathbf{T} = \mathbf{I} - \mathbf{r} \otimes \mathbf{r} \qquad . \tag{E3}$$

The symmetric transformation matrix (tensor) $\mathbf{T}$ has a simple zero eigenvalue, $\lambda = 0$, with the corresponding eigenvector $\mathbf{r}$, and a double eigenvalue, $\lambda = 1$, with the eigenvectors normal to the ray.

The non-normalized vector, $\partial v_{\text{ray}} / \partial \mathbf{r}$, is just a set of the partial derivatives. Each component of this vector, $\partial v_{\text{ray}} / \partial r_i$, characterizes the ray velocity magnitude variation vs. an infinitesimal change of the specified Cartesian component of the direction vector, $r_i$, where the two other components of the direction vector are kept fixed. The normalized gradient vector takes into account that a single component of the direction vector cannot change without corresponding adjustment of the two other components, such that a) the direction update caused by the change of the given component $\delta r_i$ is not affected by this adjustment, and b) adjustment of the two other components keeps the updated direction normalized to the unit length, $|\mathbf{r}| = 1$, $|\mathbf{r} + \delta \mathbf{r}| = 1$.

Combining equations E2 and E3, we obtain,

$$\nabla_{\mathbf{r}} v_{\text{ray}} = -v_{\text{ray}}^2 \mathbf{T} \mathbf{p} = -v_{\text{ray}}^2 \mathbf{p} + v_{\text{ray}}^2 (\mathbf{p} \cdot \mathbf{r}) \mathbf{r} = v_{\text{ray}} \mathbf{r} - v_{\text{ray}}^2 \mathbf{p} = \mathbf{v}_{\text{ray}} - v_{\text{ray}}^2 \mathbf{p} \qquad . \tag{E4}$$

The directional gradient of the ray velocity is normal to the ray (equation A10). This means that in the expression $\mathbf{v}_{\text{ray}} - v_{\text{ray}}^2 \mathbf{p}$, the second term fully cancels the first term (the ray velocity vector), and has also the component normal to the ray. This leads to the resulting formula for the directional gradient of the ray velocity,

$$\nabla_{\mathbf{r}} v_{\text{ray}} = -v_{\text{ray}}^2 \, \mathbf{r} \times \mathbf{p} \times \mathbf{r} = -\mathbf{v}_{\text{ray}} \times \mathbf{p} \times \mathbf{v}_{\text{ray}} \qquad . \qquad \text{(E5)}$$

Directional Hessian of the ray velocity

The directional Hessian of the ray velocity magnitude is also computed in two stages, obtaining first the non-normalized and then the normalized objects. The non-normalized directional Hessian reads,

$$\frac{\partial^2 v_{\text{ray}}}{\partial \mathbf{r}^2} = \frac{2}{v_{\text{ray}}} \frac{\partial v_{\text{ray}}}{\partial \mathbf{r}} \otimes \frac{\partial v_{\text{ray}}}{\partial \mathbf{r}} - v_{\text{ray}}^2 \frac{\partial \mathbf{p}}{\partial \mathbf{r}} \qquad . \qquad \text{(E6)}$$

Thus, in order to find the non-normalized directional Hessian of the ray velocity, $\partial^2 v_{\text{ray}} / \partial \mathbf{r}^2$, we need first to establish the (non-normalized) directional gradient of the slowness vector, $\mathbf{p}_{\mathbf{r}} = \partial \mathbf{p} / \partial \mathbf{r}$, which is a tensor. To compute the slowness gradient $\mathbf{p}_{\mathbf{r}}$, consider the target function gradient, $f_{\mathbf{p}}$, defined in equation D4, where all three components $f_{\mathbf{p},k}$ vanish, $k = 1,2,3$. This set consists of three nonlinear equations, $f_{\mathbf{p},k} = 0$, three unknown variables (slowness components), $p_1, p_2, p_3$, and three known parameters, $r_1, r_2, r_3$ (ray velocity direction components). The three functions, $f_{\mathbf{p},k}$, can be considered a vector function, $\mathbf{f}(\mathbf{x},\mathbf{p},\mathbf{r})$, where $\mathbf{f} \equiv f_{\mathbf{p}}$.

Assume that the nonlinear equation set D4 has already been solved, and all components of the slowness vector $p_i$ have been found for the given values of the ray velocity components $r_i$ , $i = 1,2,3$. The goal is to find the derivatives of each slowness component wrt each ray direction component, $\partial p_i / \partial r_j$ , $i, j = 1,2,3$. We can even consider a more general case when there are $n$ equations and variables, and $m$ parameters. In our case $n = m = 3$, but these two numbers may be generally different. Assume that one of the parameters, $r_i$, obtains an infinitesimal variation, while all other parameters remain fixed. For this change in the value of $r_i$, equation set D4 still holds, i.e. all functions $f_k$ accept the same value – zero. Since the functions $f_k$ are constant (zeros), this means that the full derivative of these functions wrt $r_i$ vanishes. These functions (the left-hand sides of the nonlinear equation set D4, whose right sides are zero) can be presented as,

$$f_k \left[ r_1, r_2, r_3, p_1(r_1, r_2, r_3), p_2(r_1, r_2, r_3), p_3(r_1, r_2, r_3) \right] = 0 \qquad . \qquad \text{(E7)}$$

The full derivative of the functions $f_k$ wrt parameter $r_i$ reads (e.g., Courant, 2010),

$$\frac{df_k}{dr_i} = \frac{\partial f_k}{\partial r_i} + \frac{\partial f_k}{\partial p_1}\frac{\partial p_1}{\partial r_i} + \frac{\partial f_k}{\partial p_2}\frac{\partial p_2}{\partial r_i} + \frac{\partial f_k}{\partial p_3}\frac{\partial p_3}{\partial r_i} = 0, \qquad k = 1, 2, 3 \qquad . \qquad \text{(E8)}$$

Since there are three functions $f_k$, or $n$ functions in general, we obtain a linear set of $n$ equations and $n$ unknowns. The unknown values in equation E8 are $\begin{bmatrix} \frac{\partial p_1}{\partial r_i} & \frac{\partial p_2}{\partial r_i} & \frac{\partial p_3}{\partial r_i} \end{bmatrix}$, where index $i$ is fixed. Term $\partial f_k / \partial r_i$ is moved to the right side of the equation set, and we rearrange equation E8 in a matrix form, with a square matrix of dimension $n$, in our case $n = 3$,

$$\begin{bmatrix} \frac{\partial f_1}{\partial p_1} & \frac{\partial f_1}{\partial p_2} & \frac{\partial f_1}{\partial p_3} \\ \frac{\partial f_2}{\partial p_1} & \frac{\partial f_2}{\partial p_2} & \frac{\partial f_2}{\partial p_3} \\ \frac{\partial f_3}{\partial p_1} & \frac{\partial f_3}{\partial p_2} & \frac{\partial f_3}{\partial p_3} \end{bmatrix} \begin{bmatrix} \frac{\partial p_1}{\partial r_i} \\ \frac{\partial p_2}{\partial r_i} \\ \frac{\partial p_3}{\partial r_i} \end{bmatrix} = - \begin{bmatrix} \frac{\partial f_1}{\partial r_i} \\ \frac{\partial f_2}{\partial r_i} \\ \frac{\partial f_3}{\partial r_i} \end{bmatrix} \quad , \qquad i = 1, 2, 3 \qquad . \qquad \text{(E9)}$$

Here $i=1,2,3$ means that there are three such sets, or in general $m$ such sets. Each solution of a linear set delivers a column,

$$\begin{bmatrix} \dfrac{\partial p_1}{\partial r_i} & \dfrac{\partial p_2}{\partial r_i} & \dfrac{\partial p_3}{\partial r_i} \end{bmatrix}^T \qquad , \qquad \text{(E10)}$$

i.e., derivatives of all variables $p_k$ wrt a single parameter $r_i$ . This column can be arranged as,

$$\begin{bmatrix} \dfrac{\partial p_1}{\partial r_i} \\ \dfrac{\partial p_2}{\partial r_i} \\ \dfrac{\partial p_3}{\partial r_i} \end{bmatrix} = - \begin{bmatrix} \dfrac{\partial f_1}{\partial p_1} & \dfrac{\partial f_1}{\partial p_2} & \dfrac{\partial f_1}{\partial p_3} \\ \dfrac{\partial f_2}{\partial p_1} & \dfrac{\partial f_2}{\partial p_2} & \dfrac{\partial f_2}{\partial p_3} \\ \dfrac{\partial f_3}{\partial p_1} & \dfrac{\partial f_3}{\partial p_2} & \dfrac{\partial f_3}{\partial p_3} \end{bmatrix}^{-1} \begin{bmatrix} \dfrac{\partial f_1}{\partial r_i} \\ \dfrac{\partial f_2}{\partial r_i} \\ \dfrac{\partial f_3}{\partial r_i} \end{bmatrix} \quad , \qquad i=1,2,3 \qquad . \qquad \text{(E11)}$$

In a similar way we obtain the other columns, i.e., the derivatives wrt the other parameters. We arrange equation E11 for the derivatives of all variables wrt all parameters,

$$\underbrace{\begin{bmatrix} \dfrac{\partial p_1}{\partial r_1} & \dfrac{\partial p_1}{\partial r_2} & \dfrac{\partial p_1}{\partial r_3} \\ \dfrac{\partial p_2}{\partial r_1} & \dfrac{\partial p_2}{\partial r_2} & \dfrac{\partial p_2}{\partial r_3} \\ \dfrac{\partial p_3}{\partial r_1} & \dfrac{\partial p_3}{\partial r_2} & \dfrac{\partial p_3}{\partial r_3} \end{bmatrix}}_{\text{matrix } n \times m} = - \underbrace{\begin{bmatrix} \dfrac{\partial f_1}{\partial p_1} & \dfrac{\partial f_1}{\partial p_2} & \dfrac{\partial f_1}{\partial p_3} \\ \dfrac{\partial f_2}{\partial p_1} & \dfrac{\partial f_2}{\partial p_2} & \dfrac{\partial f_2}{\partial p_3} \\ \dfrac{\partial f_3}{\partial p_1} & \dfrac{\partial f_3}{\partial p_2} & \dfrac{\partial f_3}{\partial p_3} \end{bmatrix}^{-1}}_{\text{matrix } n \times n} \underbrace{\begin{bmatrix} \dfrac{\partial f_1}{\partial r_1} & \dfrac{\partial f_1}{\partial r_2} & \dfrac{\partial f_1}{\partial r_3} \\ \dfrac{\partial f_2}{\partial r_1} & \dfrac{\partial f_2}{\partial r_2} & \dfrac{\partial f_2}{\partial r_3} \\ \dfrac{\partial f_3}{\partial r_1} & \dfrac{\partial f_3}{\partial r_2} & \dfrac{\partial f_3}{\partial r_3} \end{bmatrix}}_{\text{matrix } n \times m} \qquad . \qquad \text{(E12)}$$

As mentioned, in our case, $m=n=3$ (three components of the slowness vector and three components of the ray velocity direction). In a shorthand notation, equation E12 reads,

$$\mathbf{p_r} = -\mathbf{f_p}^{-1}\,\mathbf{f_r} \quad \text{or} \quad \mathbf{p_r} = -f_{\mathbf{pp}}^{-1}\, f_{\mathbf{pr}} \qquad , \qquad \text{(E13)}$$

where matrix $f_{\mathbf{pp}}$ is regular (invertible), while $f_{\mathbf{pr}}$ is singular. Matrices $f_{\mathbf{pp}}$ and $f_{\mathbf{pr}}$ are listed in equations D9 and D10, respectively. Note that since $\mathbf{p} = \mathbf{L_r}$, then $\mathbf{p_x} = \mathbf{L_{rx}}$, $\mathbf{p_r} = \mathbf{L_{rr}}$, and $L_{\mathbf{rr}}$ is also a singular matrix (and positive semidefinite). The non-normalized directional gradient of the slowness vector, $\mathbf{p_r}$, is symmetric and normal to the ray, $\mathbf{p_r}\,\mathbf{r} = 0$. This tensor is then used to obtain the non-normalized directional Hessian of the ray velocity in equation E6.

Recall that the matrix $\partial^2 v_{\text{ray}} / \partial \mathbf{r}^2$ on the left-hand side of equation E6 is non-normalized and will be subjected to the normalization. The normalization of the Hessians differs from the normalization of the gradient. To obtain the normalized Hessian, we need not only the non-normalized Hessian, but also the non-normalized gradient, $\partial v_{\text{ray}} / \partial \mathbf{r}$. Still, it is a linear operator, and the components of the matrix in this operator depend on the ray direction alone,

$$\nabla_{\mathbf{r}} \nabla_{\mathbf{r}} v_{\text{ray}} = \mathbf{E} \frac{\partial v_{\text{ray}}}{\partial \mathbf{r}} + \mathbf{T} \frac{\partial^2 v_{\text{ray}}}{\partial \mathbf{r}^2} \mathbf{T} \qquad , \qquad \text{(E14)}$$

where $\mathbf{T}$ is the second-order symmetric transformation tensor defined in equation E3, and $\mathbf{E}$ is a third-order super-symmetric tensor (i.e., its three indices can be swapped in any order), defined as,

$$\mathbf{E} = -\mathbf{T} \otimes \mathbf{r} - \mathbf{r} \otimes \mathbf{T} - \left(\mathbf{T} \otimes \mathbf{r}\right)^{T\{1,3,2\}} \qquad . \qquad \text{(E15)}$$

Remark: The linear operator normalizing the directional Hessian in equation E14, along with the definition of its gradient-related third-order tensor $\mathbf{E}$ in equation E15, are equivalent to those of equations 42-47 in Ravve and Koren (2019), but this operator is presented here in a compact form, involving only physical vectors and tensors.

Taking into account that $\mathbf{T} = \mathbf{I} - \mathbf{r} \otimes \mathbf{r}$ (see the second equation of set E3), equation E15 simplifies to,

$$\mathbf{E} = 3\mathbf{r} \otimes \mathbf{r} \otimes \mathbf{r} - \mathbf{I} \otimes \mathbf{r} - \mathbf{r} \otimes \mathbf{I} - \left(\mathbf{I} \otimes \mathbf{r}\right)^{T\{1,3,2\}} \quad , \tag{E16}$$

where $\mathbf{I}$ is the second-order identity tensor (matrix), resulting in,

$$E_{ijk} = 3 r_i\, r_j\, r_k - \delta_{ij}\, r_k - \delta_{jk}\, r_i - \delta_{ki}\, r_j \quad , \tag{E17}$$

where $\delta_{lm}$ is the Kronecker delta.

Due to its symmetry (obvious from equation E17), tensor $\mathbf{E}$ has only ten distinct components (out of the twenty-seven), defined, for example, by three orientation parameters (e.g., Euler's angles) and seven rotational invariants (e.g., the non-negative eigenvalues). Several types of eigenvalues exist for a third-order tensor. We use the Z-eigenvalues, $\lambda$, and their corresponding eigenvectors, $\mathbf{v}$, defined by Qi (2005),

$$\mathbf{E}\mathbf{v}\mathbf{v} = \lambda \mathbf{v} \ , \quad \mathbf{v} \cdot \mathbf{v} = 1 \ , \quad \lambda \geq 0 \quad , \tag{E18}$$

where $\mathbf{E}\mathbf{v}$ is a second-order tensor (matrix), and $\mathbf{E}\mathbf{v}\mathbf{v}$ is a vector. Multiplying this vector in the first equation of set E18 scalarly by either $\mathbf{v}$ or $\mathbf{r}$ and introducing tensor $\mathbf{E}$ from equation E15, we obtain a set of two equations,

$$\left(\mathbf{v} \cdot \mathbf{r}\right)^3 - \mathbf{v} \cdot \mathbf{r} = \lambda / 3 \ , \quad \left(\mathbf{v} \cdot \mathbf{r}\right)^2 - \lambda \mathbf{v} \cdot \mathbf{r} = 1 \ , \quad \lambda \geq 0 \quad . \tag{E19}$$

Set E19 has two solutions corresponding to a simple zero eigenvalue, $\lambda = 0$, $\mathbf{v} = \pm\mathbf{r}$, and a nonzero eigenvalue of the algebraic multiplicity six, $\lambda = 2/\sqrt{3}$, $\mathbf{v}\cdot\mathbf{r} = -1/\sqrt{3}$. The eigenvectors of the multiple eigenvalue belong to a conic surface with the axis $-\mathbf{r}$.

The first term on the right-hand side of equation E14, leads to a symmetric matrix,

$$\begin{aligned}\mathbf{E}\frac{\partial v_{\text{ray}}}{\partial \mathbf{r}} &= -v_{\text{ray}}^2 \mathbf{E}\mathbf{p} = -v_{\text{ray}}^2\left[3\mathbf{r}\otimes\mathbf{r}(\mathbf{p}\cdot\mathbf{r}) - \mathbf{I}(\mathbf{p}\cdot\mathbf{r}) - \mathbf{p}\otimes\mathbf{r} - \mathbf{r}\otimes\mathbf{p}\right] \\ &= v_{\text{ray}}^2(\mathbf{p}\otimes\mathbf{r} + \mathbf{r}\otimes\mathbf{p}) - 2v_{\text{ray}}\mathbf{r}\otimes\mathbf{r} + v_{\text{ray}}\mathbf{T} \quad .\end{aligned} \tag{E20}$$

The second term on the right-hand side of equation E14, with the use of equations E2 and E6, simplifies to,

$$\begin{aligned}&\mathbf{T}\frac{\partial^2 v_{\text{ray}}}{\partial \mathbf{r}^2}\mathbf{T} = \mathbf{T}\left(\frac{2}{v_{\text{ray}}}\frac{\partial v_{\text{ray}}}{\partial \mathbf{r}}\otimes\frac{\partial v_{\text{ray}}}{\partial \mathbf{r}} - v_{\text{ray}}^2\frac{\partial \mathbf{p}}{\partial \mathbf{r}}\right)\mathbf{T} = 2v_{\text{ray}}^3\mathbf{T}(\mathbf{p}\otimes\mathbf{p})\mathbf{T} - v_{\text{ray}}^2\mathbf{T}\frac{\partial \mathbf{p}}{\partial \mathbf{r}}\mathbf{T} = \\ &2v_{\text{ray}}^3(\mathbf{I}-\mathbf{r}\otimes\mathbf{r})\mathbf{p}\otimes\mathbf{p}(\mathbf{I}-\mathbf{r}\otimes\mathbf{r}) - v_{\text{ray}}^2(\mathbf{I}-\mathbf{r}\otimes\mathbf{r})\frac{\partial \mathbf{p}}{\partial \mathbf{r}}(\mathbf{I}-\mathbf{r}\otimes\mathbf{r}) = 2v_{\text{ray}}^3\mathbf{p}\otimes\mathbf{p} \\ &-2v_{\text{ray}}^2(\mathbf{p}\otimes\mathbf{r}+\mathbf{r}\otimes\mathbf{p}) + 2v_{\text{ray}}\mathbf{r}\otimes\mathbf{r} - v_{\text{ray}}^2\left[\frac{\partial \mathbf{p}}{\partial \mathbf{r}} - \mathbf{r}\otimes\left(\mathbf{r}\frac{\partial \mathbf{p}}{\partial \mathbf{r}}\right) - \left(\frac{\partial \mathbf{p}}{\partial \mathbf{r}}\mathbf{r}\right)\otimes\mathbf{r} + \left(\mathbf{r}\frac{\partial \mathbf{p}}{\partial \mathbf{r}}\mathbf{r}\right)\mathbf{r}\otimes\mathbf{r}\right] \ ,\end{aligned} \tag{E21}$$

which is a symmetric matrix as well. Recall that the ray direction is the eigenvector of the symmetric matrix $\partial\mathbf{p}/\partial\mathbf{r}$, with the corresponding zero eigenvalue, $\mathbf{r}(\partial\mathbf{p}/\partial\mathbf{r}) = (\partial\mathbf{p}/\partial\mathbf{r})\mathbf{r} = 0$, where $\partial\mathbf{p}/\partial\mathbf{r} = \mathbf{p}_\mathbf{r} = L_{\mathbf{rr}}$; thus only the first item in the square brackets on the right-hand side of equation E21 does not vanish. Combining equations E14, E20 and E21, we obtain the final expression for the normalized directional Hessian of the ray velocity,

$$\nabla_\mathbf{r}\nabla_\mathbf{r} v_{\text{ray}} = 2v_{\text{ray}}^3\mathbf{p}\otimes\mathbf{p} - v_{\text{ray}}^2(\mathbf{p}\otimes\mathbf{r}+\mathbf{r}\otimes\mathbf{p}) - v_{\text{ray}}\left(v_{\text{ray}}\mathbf{p}_\mathbf{r} - \mathbf{T}\right) \quad . \tag{E22}$$

For isotropic media, where $\mathbf{p} = \mathbf{r} / v_{\mathrm{ray}}$ and $\mathbf{p_r} = \mathbf{T} / v_{\mathrm{ray}}$, this directional Hessian vanishes.

<u>Spatial gradient of the ray velocity</u>

Next, we obtain the spatial gradient $\nabla_{\mathbf{x}} v_{\mathrm{ray}}$ of the ray velocity (it requires the spatial gradient of the slowness vector, $\mathbf{p_x} = \partial \mathbf{p} / \partial \mathbf{x}$),

$$\nabla_{\mathbf{x}} v_{\mathrm{ray}} = -v_{\mathrm{ray}}^2 \mathbf{p}_{\mathbf{x}}^T \mathbf{r} \qquad . \tag{E23}$$

Notice: There is a typo in equation 53 of Ravve and Koren (2019): The transpose operator for the slowness gradient matrix, $\partial \mathbf{p} / \partial \mathbf{x} = \mathbf{p_x}$, is lost there. Equation E23 is the correct one.

The spatial gradient of the slowness is geven by a relationship similar to E13,

$$\mathbf{p_x} = -\mathbf{f}_{\mathbf{p}}^{-1} \mathbf{f_x} \quad \text{or} \quad \mathbf{p_x} = -f_{\mathbf{pp}}^{-1} f_{\mathbf{px}} \qquad , \tag{E24}$$

Matrices $f_{\mathbf{pp}}$ and $f_{\mathbf{px}}$ are listed in equations D9 and D11, respectively. Note that there is a simpler workaround to compute the spatial slowness gradient directly from the Hamiltonian,

$$\mathbf{p_x} = -H_{\mathbf{pp}}^{-1} H_{\mathbf{px}} = L_{\mathbf{rx}} \qquad , \tag{E25}$$

(the relationships between the Hamiltonian's and Lagrangian's Hessians are derived in Part VI). Any type of the Hamiltonian can be applied to compute the spatial gradient of the slowness with equation E25. However, the directional gradient of the slowness cannot be computed through the Hamiltonian and its Hessains (because the Hamiltonian does not explicitly depend on the ray direction), and equation E13 should be used.

Spatial Hessian of the ray velocity

First, we compute the spatial Hessian of the slowness vector, which is the third-order tensor,

$$\mathbf{p_{xx}} = -\mathbf{f_p}^{-1}\left[\left(\mathbf{p_x}^T \, \mathbf{f_{pp}}^{T\{2,1,3\}} \, \mathbf{p_x}\right)^{T\{2,1,3\}} + \left(\mathbf{p_x}^T \, \mathbf{f_{px}}^{T\{2,1,3\}}\right)^{T\{2,1,3\}} + \mathbf{f_{xp}} \, \mathbf{p_x} + \mathbf{f_{xx}}\right] \quad , \qquad \text{(E26)}$$

where $\mathbf{f} = f_\mathbf{p}$. The spatial Hessian of the ray velocity then reads,

$$\nabla_\mathbf{x}\nabla_\mathbf{x} v_\text{ray} = \frac{2}{v_\text{ray}} \nabla_\mathbf{x} v_\text{ray} \otimes \nabla_\mathbf{x} v_\text{ray} - v_\text{ray}^2 \, \mathbf{r} \, \mathbf{p_{xx}} \quad . \qquad \text{(E27)}$$

Notice: There is a typo in equation 61b of Ravve and Koren (2019). Equation E27 is correct.

Mixed Hessians of the ray velocity

First, we compute the mixed Hessians of the slowness vector,

$$\mathbf{p_{xr}} = -\mathbf{f_p}^{-1}\left[\left(\mathbf{p_x}^T \, \mathbf{f_{pp}}^{T\{2,1,3\}} \, \mathbf{p_r}\right)^{T\{2,1,3\}} + \left(\mathbf{p_x}^T \, \mathbf{f_{pr}}^{T\{2,1,3\}}\right)^{T\{2,1,3\}} + \mathbf{f_{xp}} \, \mathbf{p_r} + \mathbf{f_{xr}}\right] \quad , \qquad \text{(E28)}$$

$$\mathbf{p_{rx}} = -\mathbf{f_p}^{-1}\left[\left(\mathbf{p_r}^T \, \mathbf{f_{pp}}^{T\{2,1,3\}} \, \mathbf{p_x}\right)^{T\{2,1,3\}} + \left(\mathbf{p_r}^T \, \mathbf{f_{px}}^{T\{2,1,3\}}\right)^{T\{2,1,3\}} + \mathbf{f_{rp}} \, \mathbf{p_x} + \mathbf{f_{rx}}\right] \quad . \qquad \text{(E29)}$$

The non-normalized mixed Hessians of the ray velocity read,

$$\frac{\partial^2 v_\text{ray}}{\partial\mathbf{x}\partial\mathbf{r}} = \frac{2}{v_\text{ray}} \frac{\partial v_\text{ray}}{\partial\mathbf{x}} \otimes \frac{\partial v_\text{ray}}{\partial\mathbf{r}} - v_\text{ray}^2 \left(\mathbf{p_x}^T + \mathbf{r} \, \mathbf{p_{xr}}\right) \quad , \qquad \text{(E30)}$$

$$\frac{\partial^2 v_\text{ray}}{\partial\mathbf{r}\partial\mathbf{x}} = \frac{2}{v_\text{ray}} \frac{\partial v_\text{ray}}{\partial\mathbf{r}} \otimes \frac{\partial v_\text{ray}}{\partial\mathbf{x}} - v_\text{ray}^2 \left(\mathbf{p_x} + \mathbf{r} \, \mathbf{p_{rx}}\right) \quad , \qquad \text{(E31)}$$

Notics: There is a typo in equation 68b of Ravve and Koren (2019). Equation E30 is correct. With the auxiliary identity,

$$\mathbf{r}\,L_{\mathbf{rr}}=0 \quad \to \quad \mathbf{r}\,\mathbf{p}_{\mathbf{r}}=0 \quad \to \quad \mathbf{r}\,\mathbf{p}_{\mathbf{xr}}=\mathbf{r}\,\mathbf{p}_{\mathbf{rx}}=0 \qquad , \tag{E32}$$

equations E30 and E31 simplify to,

$$\frac{\partial^2 v_{\text{ray}}}{\partial\mathbf{x}\partial\mathbf{r}}=\frac{2}{v_{\text{ray}}}\frac{\partial v_{\text{ray}}}{\partial\mathbf{x}}\otimes\frac{\partial v_{\text{ray}}}{\partial\mathbf{r}}-v_{\text{ray}}^2\,\mathbf{p}_{\mathbf{x}}^T \qquad , \tag{E33}$$

$$\frac{\partial^2 v_{\text{ray}}}{\partial\mathbf{r}\partial\mathbf{x}}=\frac{2}{v_{\text{ray}}}\frac{\partial v_{\text{ray}}}{\partial\mathbf{r}}\otimes\frac{\partial v_{\text{ray}}}{\partial\mathbf{x}}-v_{\text{ray}}^2\,\mathbf{p}_{\mathbf{x}} \qquad , \tag{E34}$$

The normalized mixed Hessians are,

$$\nabla_{\mathbf{x}}\nabla_{\mathbf{r}}v_{\text{ray}}=\frac{\partial^2 v_{\text{ray}}}{\partial\mathbf{x}\partial\mathbf{r}}\mathbf{T} \quad , \qquad \nabla_{\mathbf{r}}\nabla_{\mathbf{x}}v_{\text{ray}}=\mathbf{T}\frac{\partial^2 v_{\text{ray}}}{\partial\mathbf{r}\partial\mathbf{x}} \qquad . \tag{E35}$$

Notice: There is a typo in equations 69 and 70 of Ravve and Koren (2019). Equation E35 is correct.

The normalized mixed gradients of the ray velocity can be also obtained directly from equation E5, computing the spatial gradient of its left- and right-hand sides,

$$\nabla_{\mathbf{r}}\nabla_{\mathbf{x}}v_{\text{ray}}=\frac{2}{v_{\text{ray}}}\nabla_{\mathbf{r}}v_{\text{ray}}\otimes\nabla_{\mathbf{x}}v_{\text{ray}}-v_{\text{ray}}^2\left[\left(\mathbf{r}\times\mathbf{p}_{\mathbf{x}}\right)^T\times\mathbf{r}\right]^T \qquad , \tag{E36}$$

$$\nabla_{\mathbf{x}}\nabla_{\mathbf{r}}v_{\text{ray}}=\frac{2}{v_{\text{ray}}}\nabla_{\mathbf{x}}v_{\text{ray}}\otimes\nabla_{\mathbf{r}}v_{\text{ray}}-v_{\text{ray}}^2\left(\mathbf{r}\times\mathbf{p}_{\mathbf{x}}\right)^T\times\mathbf{r} \qquad , \tag{E37}$$

Note that the following identity holds,

$$\left[ \left( \mathbf{r} \times \mathbf{p_x} \right)^T \times \mathbf{r} \right]^T = \mathbf{T}\mathbf{p_x} \quad . \tag{E38}$$

With this identity, we arrive to the final simple forms,

$$\nabla_\mathbf{r} \nabla_\mathbf{x} v_{\text{ray}} = \frac{2}{v_{\text{ray}}} \nabla_\mathbf{r} v_{\text{ray}} \otimes \nabla_\mathbf{x} v_{\text{ray}} - v_{\text{ray}}^2 \mathbf{T}\mathbf{p_x} \quad , \tag{E39}$$

$$\nabla_\mathbf{x} \nabla_\mathbf{r} v_{\text{ray}} = \frac{2}{v_{\text{ray}}} \nabla_\mathbf{x} v_{\text{ray}} \otimes \nabla_\mathbf{r} v_{\text{ray}} - v_{\text{ray}}^2 \mathbf{p_x}^T \mathbf{T} \quad , \tag{E40}$$

The same results can be obtained by introducing equations E33 and E34 into E35.

Relstionship between the gradient $L_\mathbf{x}$ and the mixed Hessian $L_\mathbf{xr}$

Concluding the discussion on the spatial and directional derivatives of the ray veloicity magnitude, we derive an important relationship between the spatial gradient of the Lagrangian, $L_\mathbf{x}$, and its mixed Hessian, $L_\mathbf{xr}$. It follows from the Euler-Lagrange equation,

$$\frac{dL_\mathbf{r}}{ds} = L_\mathbf{x} \quad \to \quad \frac{d\mathbf{p}}{ds} = -\frac{\nabla_\mathbf{x} v_{\text{ray}}}{v_{\text{ray}}^2} \quad . \tag{E41}$$

With equation E23, this relationship simplifies to,

$$\frac{d\mathbf{p}}{ds} = \mathbf{p_x}^T \mathbf{r} \quad \to \quad \frac{dL_\mathbf{r}}{ds} = L_\mathbf{rx}^T \mathbf{r} \quad \to \quad L_\mathbf{x} = L_\mathbf{xr} \mathbf{r} \quad , \tag{E42}$$

and leads to,

$$L_{\mathbf{x}} = L_{\mathbf{xr}}\mathbf{r} \quad . \tag{E43}$$

On the other hand,

$$\frac{d\mathbf{p}}{ds} = \frac{\partial \mathbf{p}}{\partial \mathbf{x}}\frac{d\mathbf{x}}{ds} + \frac{\partial \mathbf{p}}{\partial \mathbf{r}}\frac{d\mathbf{r}}{ds} = \mathbf{p}_{\mathbf{x}}\mathbf{r} + \mathbf{p}_{\mathbf{r}}\dot{\mathbf{r}} \quad . \tag{E44}$$

Combining equations E42 and E44, we obtain,

$$\left(\mathbf{p}_{\mathbf{x}}^T - \mathbf{p}_{\mathbf{x}}\right)\mathbf{r} = \mathbf{p}_{\mathbf{r}}\,\dot{\mathbf{r}} \quad \rightarrow \quad \left(L_{\mathbf{xr}} - L_{\mathbf{rx}}\right)\mathbf{r} = L_{\mathbf{rr}}\,\dot{\mathbf{r}} \quad . \tag{E45}$$

This result is in agreement with the fact that in homogeneous anisotropic media, where the spatial and mixed Hessians of the Lagreangian vanish (in particular, $L_{\mathbf{rx}}$ and $L_{\mathbf{xr}}$), but the directional Hessain, $L_{\mathbf{rr}}$, exists, the ray trajectories are straight lines: The curvature vector, $\dot{\mathbf{r}}$, on the right side of equation E45 is zero. Note that the matrix in brackets (on the left-hand side) is skew-symmetric and thus, singular. The matrix on the right-hand side is singular as well.

## APPENDIX F. SPATIAL AND DIRECTIONAL DERIVATIVES OF THE LAGRANGIAN

In this appendix, using the proposed arclength-related Lagrangian (equation 2),

$$L(\mathbf{x},\mathbf{r}) = \frac{dt}{ds} = \frac{\sqrt{\mathbf{r}\cdot\mathbf{r}}}{v_{\text{ray}}(\mathbf{x},\mathbf{r})} \quad , \quad \mathbf{r} \equiv \dot{\mathbf{x}} \equiv \frac{d\mathbf{x}}{ds} \quad , \tag{F1}$$

we derive and list its first and second derivatives of the proposed Lagrangian $L[\mathbf{x}(s),\mathbf{r}(s)]$ wrt the locations and directions along the ray trajectory,

$$L_{\mathbf{x}} = \frac{\partial L}{\partial \mathbf{x}} = -\frac{\nabla_{\mathbf{x}} v_{\text{ray}} \sqrt{\mathbf{r}\cdot\mathbf{r}}}{v_{\text{ray}}^2} \quad , \quad L_{\mathbf{r}} = \frac{\partial L}{\partial \mathbf{r}} = \frac{\mathbf{r}}{v_{\text{ray}}\sqrt{\mathbf{r}\cdot\mathbf{r}}} - \frac{\nabla_{\mathbf{r}} v_{\text{ray}} \sqrt{\mathbf{r}\cdot\mathbf{r}}}{v_{\text{ray}}^2} \quad ,$$

$$L_{\mathbf{xx}} = \frac{\partial^2 L}{\partial \mathbf{x}^2} = 2\frac{\nabla_{\mathbf{x}} v_{\text{ray}} \otimes \nabla_{\mathbf{x}} v_{\text{ray}}}{v_{\text{ray}}^3} - \frac{\nabla_{\mathbf{x}}\nabla_{\mathbf{x}} v_{\text{ray}}}{v_{\text{ray}}^2} \quad ,$$

$$L_{\mathbf{xr}} = L_{\mathbf{rx}}^T = \frac{\partial^2 L}{\partial \mathbf{x} \partial \mathbf{r}} = -\frac{\nabla_{\mathbf{x}} v_{\text{ray}} \otimes \mathbf{r}}{v_{\text{ray}}^2} + 2\frac{\nabla_{\mathbf{x}} v_{\text{ray}} \otimes \nabla_{\mathbf{r}} v_{\text{ray}}}{v_{\text{ray}}^3} - \frac{\nabla_{\mathbf{x}}\nabla_{\mathbf{r}} v_{\text{ray}}}{v_{\text{ray}}^2} \quad , \tag{F2}$$

$$L_{\mathbf{rr}} = \frac{\mathbf{I} - \mathbf{r}\otimes\mathbf{r}}{v_{\text{ray}}} - \frac{\mathbf{r}\otimes\nabla_{\mathbf{r}} v_{\text{ray}} + \nabla_{\mathbf{r}} v_{\text{ray}} \otimes \mathbf{r}}{v_{\text{ray}}^2} + 2\frac{\nabla_{\mathbf{r}} v_{\text{ray}} \otimes \nabla_{\mathbf{r}} v_{\text{ray}}}{v_{\text{ray}}^3} - \frac{\nabla_{\mathbf{r}}\nabla_{\mathbf{r}} v_{\text{ray}}}{v_{\text{ray}}^2} \quad ,$$

where $L_{\mathbf{x}}, L_{\mathbf{r}}$ are vectors of length 3, and $L_{\mathbf{xx}}, L_{\mathbf{xr}}, L_{\mathbf{rx}}, L_{\mathbf{rr}}$ are square matrices (tensors) of dimension 3. Vectors $\nabla_{\mathbf{x}} v_{\text{ray}}$ and $\nabla_{\mathbf{r}} v_{\text{ray}}$ are spatial and directional gradients of the ray velocity, respectively. Tensors $\nabla_{\mathbf{x}}\nabla_{\mathbf{x}} v_{\text{ray}}$ and $\nabla_{\mathbf{r}}\nabla_{\mathbf{r}} v_{\text{ray}}$ are spatial and directional Hessians of the ray velocity and $\nabla_{\mathbf{x}}\nabla_{\mathbf{r}} v_{\text{ray}}$ and $\nabla_{\mathbf{r}}\nabla_{\mathbf{x}} v_{\text{ray}}$ are the mixed Hessians. Ravve and Koren (2019) provide a computational workflow to establish these gradients and Hessians in smooth heterogeneous general anisotropic media. The first derivatives $L_{\mathbf{x}}$ and $L_{\mathbf{r}}$ define the local traveltime gradients, while the second derivatives $L_{\mathbf{xx}}, L_{\mathbf{xr}}, L_{\mathbf{rx}}, L_{\mathbf{rr}}$ define the local traveltime Hessians, used in both kinematic and dynamic analysis. Note that $\sqrt{\mathbf{r}\cdot\mathbf{r}}$ in the formulae for the gradients, $L_{\mathbf{x}}$ and $L_{\mathbf{r}}$, is kept in order to obtain the right expressions for the Hessians. On the completion of the derivations, it can be replaced by 1.

For isotropic media, operators $\nabla_{\mathbf{r}}$ result in vanishing directional derivatives, and the gradients/Hessians of the Lagrangian simplify to,

$$L_{\mathbf{x}} = -\frac{\nabla v}{v^2} \quad , \qquad L_{\mathbf{r}} = \frac{\mathbf{r}}{v} \quad ,$$

$$L_{\mathbf{xx}} = 2\frac{\nabla v \otimes \nabla v}{v^3} - \frac{\nabla \nabla v}{v^2} \quad , \qquad L_{\mathbf{xr}} = L_{\mathbf{rx}}^T = -\frac{\nabla v \otimes \mathbf{r}}{v^2} \quad , \qquad L_{\mathbf{rr}} = \frac{\mathbf{I} - \mathbf{r} \otimes \mathbf{r}}{v} \quad , \tag{F3}$$

where all derivatives are spatial, $\nabla \equiv \nabla_{\mathbf{x}}$, and $v(\mathbf{x})$ is the medium velocity.

Remark: The Euler-Lagrange equation includes only the Lagrangian gradients and does not include its Hessians. However, for for the proposed implementation of the Eigenray kinematic method and for identifying the type of the stationary solution, the Lagrangian Hessians are also needed. They are the building blocks of the global traveltime Hessian. For the Eigenray dynamic analysis, operating with the Lagrangian Hessains is mandatory as they are the varying, arclength-dependent coefficients of the linear Jacobi equation.

## LIST OF TABLES

Table 1. Notations and definition for Hamiltonians.

Table 2. Notations and definitions for Lagrangians.

Table 1. Notations and definitions for Hamiltonians.

| # | notation | definition | flow variable | units | nickname |
|---|---|---|---|---|---|
| 1 | $H^{\varsigma}$ | any of the below | general | $[\mathrm{T}]/[\zeta]$ | general |
| 2 | $H^{\bar{\tau}}$ | $\det[\mathbf{\Gamma}-\mathbf{I}]$ | scaled time, $\bar{\tau}$ | unitless | reference |
| 3 | $H^{\tau}$ | $\dfrac{H^{\bar{\tau}}}{\mathbf{p}\cdot H_{\mathbf{p}}^{\bar{\tau}}}$ | traveltime, $\tau$ | unitless | time-related |
| 4 | $H \equiv H^{s}$ | $\dfrac{H^{\bar{\tau}}}{\sqrt{H_{\mathbf{p}}^{\bar{\tau}}\cdot H_{\mathbf{p}}^{\bar{\tau}}}}$ | arclength, $s$ | $[\mathrm{T}/\mathrm{L}]$ | arclength-related |
| 5 | $H^{\sigma}$ | $H^{\bar{\tau}}\,\dfrac{\mathbf{p}\cdot H_{\mathbf{p}}^{\bar{\tau}}}{H_{\mathbf{p}}^{\bar{\tau}}\cdot H_{\mathbf{p}}^{\bar{\tau}}}$ | parameter $\sigma$ | $\left[\mathrm{T}^2/\mathrm{L}^2\right]$ | sigma-related |
| 6 | $H^{\lambda}$ | $\dfrac{1}{2}\mathbf{g}\,\mathbf{\Gamma}\,\mathbf{g}$ | traveltime, $\tau$ | unitless | eigenvalue |
| 7 | $H^{e}$ | Part II, eq. G2 | traveltime, $\tau$ | unitless | ellipsoidal |
| 8 | $H_{\mathrm{iso}}^{(n)}$ | $\dfrac{v^{2-n}}{2}\left(\mathbf{p}\cdot\mathbf{p}-v^{-2}\right)$ | $n=0$ traveltime<br>$n=1$ arclength<br>$n=2$ sigma | unitless<br>$[\mathrm{T}/\mathrm{L}]$<br>$\left[\mathrm{T}^2/\mathrm{L}^2\right]$ | isotropic time-, arclength- or sigma-related |

Table caption: In this table, $\mathbf{\Gamma}=\mathbf{p}\cdot\tilde{\mathbf{C}}(\mathbf{x})\cdot\mathbf{p}$ is the Christoffel matrix (tensor), $\tilde{\mathbf{C}}$ is the fourth-order material stiffness tensor, $\mathbf{p}$ is the slowness vector, $\mathbf{x}$ is the position vector, $\mathbf{I}$ is the identity matrix, $\mathbf{g}$ is the normalized polarization vector. Hamiltonian $H^{\lambda}$ is used, for example, by Červený (2000, 2002a, 2002b). It equals $1/2$ along the ray, while all other Hamiltonians vanish. The time-related, arclength-related and sigma-related Hamiltonians $H^{\tau}, H^{s}$ and $H^{\sigma}$, respectively, are derived from the reference Hamiltonian $H^{\bar{\tau}}$. Hamiltonian $H^{e}$ is related to ellipsoidal orthorhombic media, and it represents a sort of an acoustic approximation (the corresponding Christoffel equation has a single root). Hamiltonian $H_{\mathrm{iso}}^{(n)}$ (Červený, 2000) can be applied for isotropic media only, where index $n$ points to the flow variable, and $v=v(\mathbf{x})$ is the isotropic velocity.

Table 2. Notations and definitions for Lagrangians.

| # | notation | definition | flow variable | units | degree | related Hamiltonian | nickname |
|---|---|---|---|---|---|---|---|
| 1 | $L^{\zeta}$ | – | general, $\zeta$ | $[\mathrm{T}]/[\zeta]$ | – | $H^{\zeta}$ | general |
| 2 | $L \equiv L^{s}$ | $\dfrac{\sqrt{\mathbf{r}\cdot\mathbf{r}}}{v_{\mathrm{ray}}(\mathbf{x},\mathbf{r})}$ | arclength, $s$ | $[\mathrm{T}/\mathrm{L}]$ | 1 | $H \equiv H^{s}$ | arclength-related |
| 3 | $\hat{L}$ | $L\dfrac{ds}{d\xi}$ | internal parameter, $\xi$ | $[\mathrm{T}]$ | 1 | – | normalized |
| 4 | $L^{U}$ | $\sqrt{\mathbf{v}_{\mathrm{ray}}\mathbf{G}\,\mathbf{v}_{\mathrm{ray}}}$ | traveltime, $\tau$ | unitless | 1 | could not be explicitly derived | unmodified |
| 5 | $L^{M}$ | $\dfrac{1}{2}\mathbf{v}_{\mathrm{ray}}\mathbf{G}\,\mathbf{v}_{\mathrm{ray}}$ | traveltime, $\tau$ | unitless | 2 | $H^{\lambda}$ | modified |

Table caption: In this table, $\mathbf{G}$ is the Finsler metric (matrix), $\dot{\mathbf{x}}_{\tau} = \mathbf{v}_{\mathrm{ray}}$ is the ray velocity vector, $v_{\mathrm{ray}}$ is its magnitude, $\mathbf{x}$ is the position along the ray, $\mathbf{r} = \dot{\mathbf{x}}$ is the normalized ray velocity direction. $L^{\xi}$ is a general Lagrangian, $L$ is our proposed Lagrangian, $L^{U}$ and $L^{M}$ are the unmodified and modified Lagrangians, respectively, suggested by Červený (2002a, 2002b). Lagrangians $L$ and $L^{M}$ are related to their corresponding Hamiltonians $H$ and $H^{\lambda}$ through the Legendre transform . The same is true for the unmodified Lagrangian $L^{U}$, but its corresponding Hamiltonian could not be explicitly derived (Červený, 2002a, page 217). "Degree" means the homogeneity degree of Lagrangian $L$ wrt the components of vector $\mathbf{r}$, and the homogeneity degree of Lagrangians $L^{U}$ and $L^{M}$ wrt the components of vector $\dot{\mathbf{x}}_{\tau} = \mathbf{v}_{\mathrm{ray}}$. Lagrangian $\hat{L}$ is used in the finite-element implementation (Part III), with the internal flow variable defined within a single finite element, $-1 \le \xi \le +1$, related to the arclength by a positive scalar metric, $ds/d\xi$.

## LIST OF FIGURES

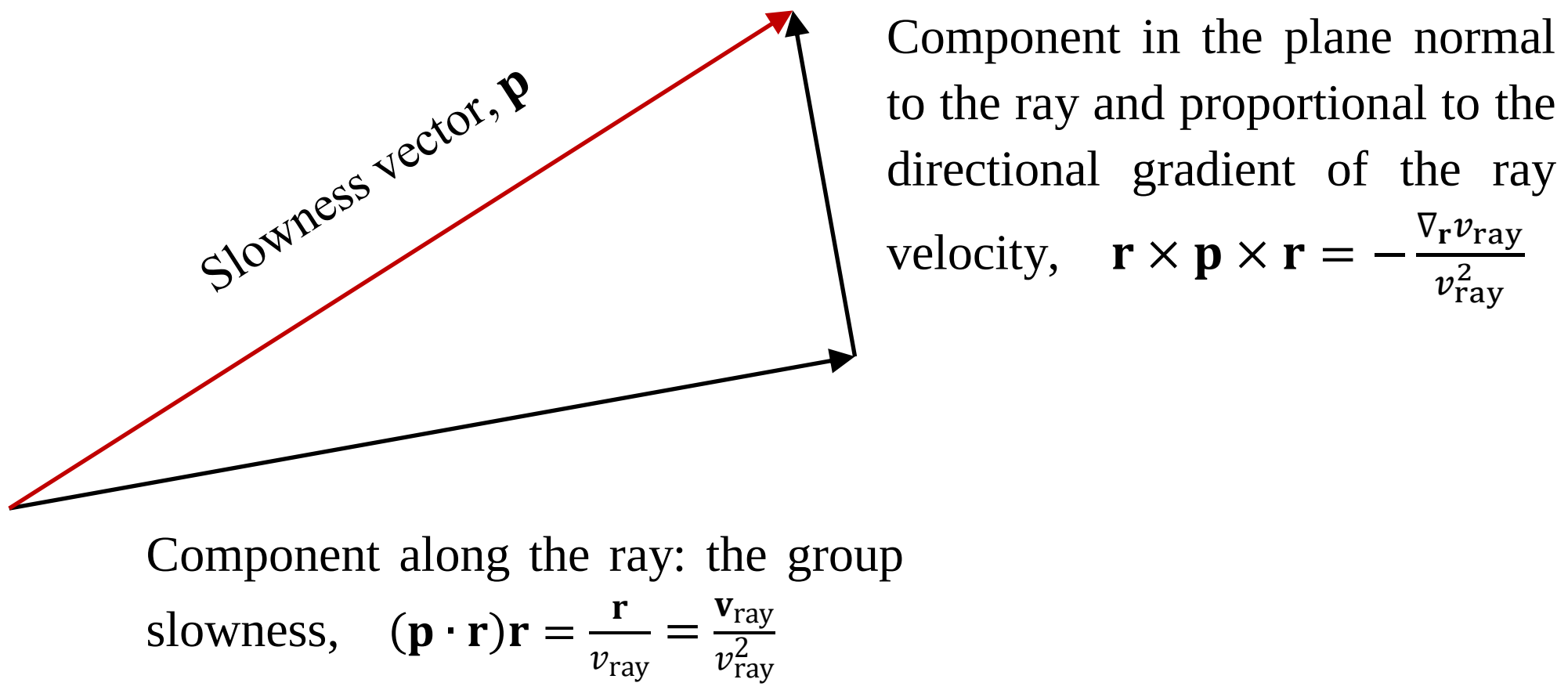

Figure 1. Decomposition of the slowness vector into components along the ray and in the normal plane.